\documentclass[draftclsnofoot,twoside,onecolumn,letter,12pt]{IEEEtran}

\usepackage{graphicx}
\usepackage{amssymb}
\usepackage{amsmath}
\usepackage{cite}
\usepackage{stfloats}

\newtheorem{definition}{Definition}
\newtheorem{theorem}{Theorem}
\newtheorem{proposition}{Proposition}

\DeclareMathOperator{\prob}{\mathbb{P}}

\newcommand{\PDF}[2]{f_{{#1}}\left({#2}\right)}
\newcommand{\CDF}[2]{F_{{#1}}\left({#2}\right)}

\newcommand{\Prob}[1]{\prob\left\{{#1}\right\}}

\newcommand{\ExpRV}[1]{\mathcal{E}\left(#1\right)}

\begin{document}

\title{
 \hspace{4cm}\\[-0.7cm]
       Outage Probability Analysis of Dual Hop Relay Networks in Presence of Interference
      }
\author{
        {
        Bappi Barua,
        Mehran Abolhasan,
        Daniel Franklin
        and
        Farzad Safaei
        }

\thanks{Bappi Barua, Mehran Abolhasan and Daniel Franklin are with the Faculty of Engineering and IT,
        University of Technology Sydney, 15 Broadway, Ultimo, NSW 2007, Australia
            (e-mail: bappi.barua@student.uts.edu.au, mehran.abolhasan@uts.edu.au, daniel.franklin@uts.edu.au).}
\thanks{Farzad Safaei is with the Faculty of Informatics, University of Wollongong, NSW 2522, Australia
            (e-mail: farzad@uow.edu.au).}

\thanks{This research was supported by the Australian Research Council (ARC) discovery research grant No. DP0879507.
            }
}

\markboth{}
        {Barua \textit{\MakeLowercase{et al.}}:
        Outage Probability Analysis of Dual Hop Relay Networks in Presence of Interference}

\maketitle



\begin{abstract}
Cooperative relaying improves the performance of wireless networks by forming a network of multiple independent virtual sources transmitting the same information as the source node. However, interference induced in the network reduces the performance of cooperative communications. In this work the statistical properties, the cumulative distribution function (CDF) and the probability density function (PDF) for a basic dual hop cooperative relay network with an arbitrary number of interferers over Rayleigh fading channels are derived. Two system models are considered: in the first system model, the interferers are only at the relay node; and in the second system model, interferers are both at the relay and the destination. This work is further extended to Nakagami-$m$ faded interfering channels. Simulation results are presented on outage probability performance to verify the theoretical analysis.
\end{abstract}


\section{Introduction}  \label{sec:A}

\PARstart{W}{ireless} channels often suffer severe performance degradation due to multipath fading and interference. The increase of spectral reuse in wireless systems exposes the network to ever-large number of interfering nodes. Ad-hoc and multiuser networks often face interference from other user nodes operating in multiple parallel channels. Interference can also appear in the network as cochannel interference. Cooperative relay communications, originally introduced by van der Meulen \cite{Meu:71:AAP} has been extensively studied in \cite{LTW:04:IT, HA:03:WCOM, HA:04:COM, BHS:08:CL} without the presence of interference. In cooperative relay networks, cooperating nodes help the information source node by jointly forming a virtual antenna array and thus providing higher diversity gain and information reception reliability. In addition, cooperating relays can offer a very flexible extension of network coverage without substantial development of infrastructure. However to date, research efforts on the interference channels have mainly concentrated on information theoretic analysis \cite{AC:75:IT, AC:78:IT, RB:79:IT, HK:81:IT, MC:85:IT, AV:11:IT} and using assisted relay networks in \cite{SGN:10:CL, QDZQA:11:SPL, SMY:12:TVT, IA:10:GLOBCOM, CDS:11:IET, XZZ:11:CL, LL:11:TVT, CY:11:CL, CZLY:10:VTC, ZJW:10:TC, CDG:11:CL, MDFT:11:IT, RDH:06:IZS, MDFT:08:CCC, TN:12:IT}.
Statistical modeling for single hop multiple antenna systems with active interferers has been performed for maximal ratio combining (MRC) receivers in \cite{ZB:05:CL, ZB:06:CL, CZF:04:CL, CVA:03:TWC, AZ:01:TVT} and using optimum combining receivers in \cite{JW:84:JSAC, SH:98:TC, VZ:00:CL, CWZ:03:TC}.

Interference in cooperative relay networks has been extensively studied in previous publications \cite{SGN:10:CL, QDZQA:11:SPL, SMY:12:TVT, IA:10:GLOBCOM, CDS:11:IET, XZZ:11:CL, LL:11:TVT, CY:11:CL, CZLY:10:VTC, ZJW:10:TC, CDG:11:CL}. Authors of \cite{SGN:10:CL} and \cite{QDZQA:11:SPL} consider a scenario where only the relay node experiences interference in interference limited networks and the receiver node remains interference free. The relays here can estimate the instantaneous CSI of interfering channels to scale the gain. A fixed gain AF relay with interference limited destination is considered in \cite{SMY:12:TVT}. However, the assumption that the AF relay gain parameter includes the instantaneous or average channel information of interfering channels (as assumed in \cite{SGN:10:CL, SMY:12:TVT, QDZQA:11:SPL, IA:10:GLOBCOM, XZZ:11:CL, CY:11:CL, CZLY:10:VTC}) requires additional computational capability at the relaying node, and in certain cases where the interfering signals are not known to the relay a priori, the technique can not be applied. Furthermore, in many previous results authors consider interference limited cooperative networks \cite{XZZ:11:CL, LL:11:TVT, CY:11:CL, CZLY:10:VTC, ZJW:10:TC, CDG:11:CL}. Outage performance of a dual hop network has been studied using a fixed gain relay in \cite{XZZ:11:CL} and hypothetical gain AF relay in \cite{LL:11:TVT} with an arbitrary number of interferers. The system is assumed to be an interference dominated network where noise power is negligible compared to interfering signal power. Closed form PDF and CDF of signal to interference ration (SIR) are derived. Avoiding the noise in receiver systems usually provides an opportunity to express some of the integrals used to obtain the CDF and the PDF in very standard formats. Outage probability using DF relays in Nakagami-$m$ fading channels is studied in \cite{CDG:11:CL}. The authors consider that the destination faces a negligible amount of interference, but non-negligible noise. The PDF and CDF of corresponding receiver SNR is derived.
In \cite{MDFT:08:CCC}, the authors have studied transmission techniques designed to manage interference using an information-theoretic approach. Bit error rate (BER) performance has been presented in \cite{LJJ:11:TWC} for a multi-user network in which multiple sources communicate with one receiver node via a common relay in a dual hop network. A different approach has also been proposed to cancel the interference at the relay node. In \cite{VL:08:TWC}, the authors have investigated the effect of cooperation in an interference-limited system utilizing a decode-and-forward (DF) relaying protocol over Rayleigh fading channels. It divides the cooperating nodes into different cooperating regions and quantifies the relation between cooperative region radius and the interference level. It also analyzes the network sum rate optimization problem for a cooperative region radius.

In this work, we derive the closed form expressions of statistical properties, the CDF and the PDF of signal to noise plus interference ratio (SINR) of an interference relay network that is operating under the influence of arbitrary number of interferes and noise. An amplify-and-forward (AF) relaying protocol is adopted due to its simplicity and ease of deployment at relay nodes. Lastly numerical results on the outage probability performance are presented to verify the theoretical analysis via Monte-Carlo simulations.

\section{System and Channel Model}  \label{sec:B}

Consider a single source-destination pair communicating via a single antenna relay without any direct link. We will denote source-relay and relay-destination links as S-R and R-D respectively. A half duplex AF protocol has been considered over independent and non identically distributed (i.n.i.d.) Rayleigh fading channels. In this paper, two different system models are investigated: system model 1 (SM 1), in which the interferers are only at the relay node; and system model 2 (SM 2), where interferers affect both the relay and destination nodes. In both models, all the interfering channels are i.n.i.d. Rayleigh faded.
The destination is assumed to have full channel state information (CSI) of the two main channels, S-R and R-D, while the relay has full CSI of the S-R channel only. The source and relay have no CSI of forwarding transmitting channels. None of the nodes, source (S), relay (R) and destination (D) possess information about the interfering channels.

The instantaneous and average signal to noise ratio (SNR) of 1st and 2nd hops are denoted as $\gamma_{i} \triangleq \frac{P \left|h_{i}\right|^2}{\sigma_{i}^2}$ and ${\lambda}_i \triangleq \frac{P \Omega_{i}}{\sigma_{i}^2}$ respectively, where $i \in \{1,2\}$, $P$ is the corresponding source and relay power; $h_{i}$ and $\Omega_{i}$ are the instantaneous and average channel gain of the $i$th hop respectively\footnote{The average channel gain, $\Omega_{i}$ is in fact the statistical average of the squared instantaneous channel gain $h_{i}$, i.e. $\Omega_{i} \triangleq \mathbb{E}\left[\left|h_{i}\right|^2\right]$.
} and
$\sigma_{i}^2$ is the one sided additive white Gaussian noise (AWGN) power at relay or destination node, i.e. $i \in \{R, D\}$.
We assume the total power of the network (source and relay) is constrained to $P_{\mathrm{tot}}$. This total power is split between the source and the relay by a power sharing coefficient $\zeta \in (0,1]$ such that the source and the relay powers are given by, $P_{S}=(1-\zeta)P_{\mathrm{tot}}$ and $P_{R}=\zeta P_{\mathrm{tot}}$ respectively. Thus if equal power sharing protocol is adopted, $\zeta$ will be 0.5.

Let there be a total of $L$ interferers in the system, and define an interferer set $\mathfrak{I}$, the set of all interfering source nodes. For example, any interferer $I_l \in \mathfrak{I}$, where $l \in \mathcal{L}$, $\mathcal{L}=\{1,2,3.....,L\}$. The elements of interfering channel row vectors $\mathbf{h}_{I,i} \in C^{L}$ represent the corresponding interference channels from the source element nodes of the interfering signal source vectors $\mathbf{x}_{I,i} \in C^{L}$, $i \in \{R, D\}$.\footnote{
$C^{L}$ denotes a $L$-dimensional complex vector.
}
Furthermore, we assume the interferer $I_l$ has individual transmit power $P_{I,l}$. Thus the instantaneous and the average interference power to noise ratio (INR) for any interferer $I_l$ is $\gamma_{I,l} \triangleq \frac{P_{I,l} \left|h_{I,l}\right|^2}{\sigma_{i}^2}$ and ${\lambda}_{I,l} \triangleq \frac{P_{I,l} \Omega_{I,l}}{\sigma_{i}^2}$ respectively. $h_{I,l}$ and $\sigma_{i}^2$ are the fading channel gain from the interfering source $I_l$ to the node $i$ and the noise power at node $i$ respectively, $l \in \mathcal{L}$ and $i \in \{R, D\}$. $\Omega_{I,l}$ is the average interfering channel gain, $\Omega_{I,l} \triangleq \mathbb{E}\left[\left|h_{l,l}\right|^2\right]$.   Throughout this paper the term INR will be used to indicate individual interferer signal power to noise ratio unless otherwise specified.

%
%
%
%
%

\subsection{System Model 1: Interference at Relay} \label{sec:B:01}

\begin{figure}[!t]
    \centerline{\includegraphics[width=0.9 \textwidth]{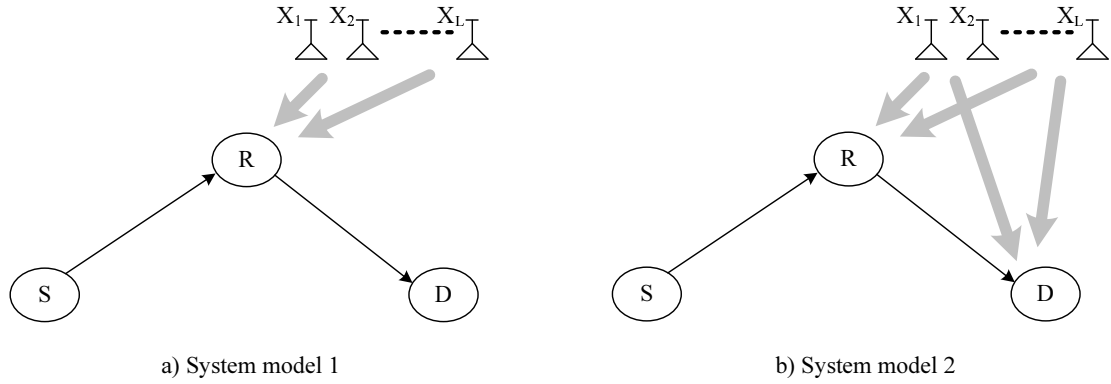}}
    \caption{
    Interference relay network.
    }
    \label{fig:A1}
\end{figure}

In the network model as shown in Fig.~\ref{fig:A1}, interference occurs only at the relay node. The received signal at the relay node $R$ is
\begin{align} \label{EEIRN:B:01:01}
y_{R}  = h_{1} x_{s} + \mathbf{h}_{I,1}\mathbf{x}_{I,1}^{T}  + n_{1}
\end{align}
where, $\mathbf{h}_{I,1} \in C^{L}$ and  $\mathbf{x}_{I,1} \in C^{L}$ are the row vectors for the interference channels and the corresponding interfering signal sources respectively. The notation $(.)^T$ represents the transpose operation on a vector. The signal at the destination node is
\begin{align}  \label{IRN:B:01:02}
y_{D} = G h_{2} h_{1} x_{s} + G h_{2} \mathbf{h}_{I,1}\mathbf{x}_{I,1}^{T}  + G h_{2} n_{1} +n_{2}
\end{align}
where $G$ is the AF relay gain. $n_1 \sim \mathcal{CN}\left(0,\sigma_1^2\right) $ and $n_2 \sim \mathcal{CN}\left(0,\sigma_2^2\right)$ are additive white Gaussian noise at the relay and destination
respectively.\footnote{$\mathcal{CN}\left(\lambda,\sigma^2\right)$ denotes a circularly symmetric complex Gaussian random variable
with mean $\lambda$ and variance $\sigma^2$.
}
At the receiver node, the signal to interference plus noise ratio (SINR) will be
\begin{align} \label{IRN:B:01:03}
\gamma_{\mathrm{SINR_1}}
    = \frac{
        {G^2} {\left|h_1 \right|^2} {\left|h_2 \right|^2} P_S
        }
        {
        {G^2} {{\left|h_2 \right|^2} \mathbf{h}_{I,1} \mathbf{\Sigma}_{I,1} \mathbf{h}_{I,1}^{\dag}}
        + {G^2} {\left|h_2 \right|^2} \sigma_{1}^2
        + \sigma_{2}^2
 }
\end{align}
where the diagonal matrix $\mathbf{\Sigma}_{I}=E \left\{\mathbf{x}_{I,1}^{\dag} \mathbf{x}_{I,1}\right\}$ is composed of interference signal powers.
With hypothetical AF relay gain $G^2=\frac{P_R}{P_S  \left|h_{1}\right|^2}$, the end-to-end SINR is \cite{HA:03:WCOM}\footnote{
A CSI assisted relay gain $G^2=\frac{P_R}{\left|h_{1}\right|^2 P_S + \sigma_{1}^2}$ proposed in \cite{LTW:04:IT} render the end-to-end SINR as
        $ \gamma_{\mathrm{SINR}}
        = \frac{
        \gamma_{1} \gamma_{2}
        }
        {
        \gamma_{1}+\gamma_{2}+\gamma_{2} \gamma_{I}+1
        }
        $.
Eq.~\eqref{IRN:B:01:04} proposes a tight upper bound on the CSI assisted SINR  $ \gamma_{\mathrm{SINR}}$.
},
\begin{align} \label{IRN:B:01:04}
\gamma_{\mathrm{SINR_{1}}}
    = \frac{
        \gamma_{1} \gamma_{2}
        }
        {
        \gamma_{1}+\gamma_{2}+\gamma_{2} \gamma_{I}
        }
\end{align}
where $\gamma_{1}$ and $\gamma_{2}$ are the instantaneous SNRs for S-R and R-D links respectively. $\gamma_{I}$ is the \emph{total} interference to noise ratio at the relay node, which is literally the sum of all individual INRs.

Note that due to the Rayleigh fading assumption, the first and second hop SNR $\gamma_1$ and $\gamma_2$ are exponentially distributed with mean parameter $\lambda_1$ and $\lambda_2$ respectively.
Later, it will be shown that under an equal power sharing protocol at the source and relay nodes, the system performance metric of system model 1 does not vary if the interferers are switched from the relay to the destination node.

\subsection{System Model 2: Interference at Relay and Destination}  \label{sec:B:02}
In this network model, interference occurs both at the relay and at the destination nodes. We let there are $L_1$ interferers at the relay node and $L_2$ interferers at the destination. All the interfering channels are independent and non-identically distributed. In this case, the received signal at the destination is
\begin{align}  \label{IRN:B:02:02}
    y_{D}
        =
        G h_{2} h_{1} x_{s}
        + G h_{2} \mathbf{h}_{I,1}\mathbf{x}_{I,1}^{T}
        + \mathbf{h}_{I,2}\mathbf{x}_{I,2}^{T}
        + G h_{2} n_{1} +n_{2}
\end{align}
where $\mathbf{x}_{I,1} \in C^{L_1}$ and $\mathbf{x}_{I,2} \in C^{L_2}$ are the vectors with interference sources for the relay and receiver nodes respectively, and $\mathbf{h}_{I,1} \in C^{L_1}$ and $\mathbf{h}_{I,2} \in C^{L_2}$ are the corresponding fading channels from interferers to the relay and receiver nodes respectively. Again, as for SM 1, $G$ is the AF relay gain, $n_1 \sim \mathcal{CN}\left(0,\sigma_1^2\right) $ and $n_2 \sim \mathcal{CN}\left(0,\sigma_2^2\right)$ are AWGN at the relay and the destination respectively.
Thus the SINR with arbitrary relay gain is given by
\begin{align} \label{IRN:B:02:03}
\gamma_{\mathrm{SINR_2}}
    = \frac{
        {G^2} {\left|h_1 \right|^2} {\left|h_2 \right|^2} P_S
        }
        {
        {G^2} {{\left|h_2 \right|^2} \mathbf{h}_{I,1} \mathbf{\Sigma}_{I,1} \mathbf{h}_{I,1}^{\dag}}
        +\mathbf{h}_{I,2} \mathbf{\Sigma}_{I,2} \mathbf{h}_{I,2}^{\dag}
        + {G^2} {\left|h_2 \right|^2} \sigma_{1}^2
        + \sigma_{2}^2
 }
\end{align}
$\mathbf{\Sigma}_{I,1}=E \left\{\mathbf{x}_{I,1}^{\dag} \mathbf{x}_{I,1}\right\}$ and $\mathbf{\Sigma}_{I,2}=E \left\{\mathbf{x}_{I,2}^{\dag} \mathbf{x}_{I,2}\right\}$ are diagonal matrices of the transmission powers of interfering signals at the relay node and the destination respectively. Applying the similar hypothetical relay gain as used in SM 1 at the AF relay, the receiver SINR is given by
\begin{align} \label{IRN:B:02:04}
\gamma_{\mathrm{SINR}_2}
    = \frac{
        \gamma_{1} \gamma_{2}
        }
        {
        \gamma_{1}+\gamma_{2}+\gamma_{1} \gamma_{I,2}+\gamma_{2} \gamma_{I,1}
        }
\end{align}
where $\gamma_{I,1}$ and $\gamma_{I,2}$ are the total interference to noise ratio at the relay node and the destination respectively.


\section{Statistical Analysis}  \label{sec:SINR:01}
In the following sections statistical theorems related to the network of  interest are developed.

\begin{definition}  \label{def:01:SINR:MD01}

Let random variables $X$, $Y$ and $U$ be statistically independent, where $X \sim \ExpRV{\frac{1}{\lambda_x}}$ and $Y \sim \ExpRV{\frac{1}{\lambda_y}}.$\footnote{
        $\ExpRV{\frac{1}{\lambda_x}}$ denotes an exponential distribution
        with a hazard rate ${\frac{1}{\lambda_x}}$.
        }
Random variable $U=\sum\limits_{l=1}^{L}U_l +1 $, where $U_l \sim \ExpRV{\frac{1}{\lambda_{u_l}}}$ are i.n.i.d. exponentially distributed random variables and $U$ has the PDF,
$
\PDF{U}{u} =
          \sum\limits_{i = 1}^{\zeta \left( \Lambda \right)}
          \sum\limits_{j = 1}^{{\tau _i}\left( \Lambda \right)}
          \frac{
          {{X_{i,j}}\left( \Lambda \right)
          {{\left(u-1\right)}^{j - 1}}
          }
          }
          {
          {\Gamma \left( {j} \right)}
          {\lambda_{u\langle i \rangle}^{j}}
          }
          e^{- \frac{{\left(u-1\right)}}{\lambda_{u \langle i \rangle}}},
$\footnote{
          where
          $\Lambda =\mathrm{diag}
          \left({{{\lambda}_{u_1}}, {{\lambda}_{u_2}},....,{{\lambda }_{u_L}}
          }\right)
          $.
          $
          \zeta \left({\Lambda}\right)
          $
          is the number of distinct diagonal elements of $\Lambda $ and
          ${\lambda_{u \langle 1\rangle }}>{\lambda_{u \langle 2\rangle }}
          >......>{\lambda_{u\langle \zeta \left(\Lambda\right)\rangle }}
          $ are the distinct diagonal elements in decreasing order.
          $ {\tau _i} \left( {\Lambda} \right)
          $ is the multiplicity of ${\lambda_{u \langle i\rangle}}$ and
          ${X_{i,j}}\left({\Lambda}\right)$
is the $(i,j)$th characteristic coefficient of $\Lambda$ \cite{SW:08:IT}.}
$u \geq 1$ \cite{BSW:07:WCOM}.
A r.v. $W$ is defined as
\begin{align}  \label{eq:W:01}
W=\frac{XY}{X+YU}
\end{align}

\end{definition}

\begin{theorem}[CDF and PDF] \label{thm01:cdf:pdf:SINR1}

The cumulative distribution function (CDF) and probability density function (PDF) of random variable $W$ are given respectively as

\begin{align} \label{CDF:MD01:01}
\CDF{W}{w}
&=
		1-
     	\sum\limits_{i = 1}^{\zeta \left( \Lambda \right)}
        \sum\limits_{j = 1}^{{\tau _i}\left( \Lambda \right)}
        {{X_{i,j}}\left( \Lambda \right)}
        e^{-w \left(\frac{1}{\lambda_x}
		+\frac{1}{{\lambda_y}}\right)}
		\left(\frac{{\lambda_x}}{{\lambda_x}+w {\lambda_{u{ \langle i \rangle }}}}\right)^j
\left[
        1+
        \sum \limits_{k=0}^{\infty }
		\sum \limits_{n=0}^{k+1}
        \binom{k+1}{n}
\right.
\nonumber
\\
& \times
\left.	
        \frac{w^{2k+2}}
        {\Gamma\left(j\right) \left({\lambda_x}
        {\lambda_y}\right)^{k+1}
        {k! (k+1)!}}	
        \left(\frac{{\lambda_x} {\lambda_{u{ \langle i \rangle }}}}{{\lambda_x}+w {\lambda_u}}
        \right)^{n}
		\left[
        2 {\Gamma}\left(j+n\right)
        \left\{{\ln}\left(\frac{w}{\sqrt{{\lambda_x} {\lambda_y}}}\right)
		-\psi_k
\right\}
\right.
\right.
\nonumber
\\
& +
\left.
\left.
        {G}_{3,2}^{1,3}
        \left(
        \frac{{\lambda_x} {\lambda_{u{ \langle i \rangle }}}}{{\lambda_x}+w {\lambda_{u{ \langle i \rangle }}}}
        \left|
        \begin{matrix}
        {1-j-n,1,1} \\
        {~~ 1,0 }
        \end{matrix}
        \right.
        \right)
\right]
\right]
\end{align}
%
%
\begin{align}  \label{PDF:MD01:01}
&\PDF{W}{w}
        =
        \sum\limits_{i = 1}^{\zeta \left( \Lambda \right)}
        \sum\limits_{j = 1}^{{\tau _i}\left( \Lambda \right)}
        {{X_{i,j}}\left( \Lambda \right)}
        e^{-w\left(\frac{1}{\lambda_x}+\frac{1}{\lambda_y}\right)}
        \left(\frac{\lambda_x}{\lambda_x+w\lambda_{u{\langle i \rangle}}}\right)^j
        \left[
        \frac{1}{\lambda_x}+\frac{1}{\lambda_y}
        +\frac{j \lambda_{u{\langle i \rangle}}}{\lambda_x+w\lambda_{u{\langle i \rangle}}}
        \right.
\nonumber
\\
&
        \left.
		-
        \sum \limits_{k=0}^{\infty }
		\sum \limits_{n=0}^{k+1}
        \binom{k+1}{n}
		\frac{w^{2k+1}}
             {\Gamma\left(j\right)
             \left({\lambda_x} {\lambda_y}\right)^{k+1} k! (k+1)!}
        \left(\frac{{\lambda_x} {\lambda_{u{\langle i \rangle}}}}{{\lambda_x}+w {\lambda_{u{\langle i \rangle}}}}
        \right)^{n}
		\!
        \left[
        {2\Gamma\left(j+n\right)}
        +\mathcal I{_1}\left(w,\lambda\right)
        \Bigg\{
        {2 k+2}
\right.
\right.
\nonumber
\\
&
\left.
\left.
        -\frac{w \left(\lambda_x+\lambda_y\right)}{{\lambda_x \lambda_y}}
        -\frac{\left(j+n\right){w \lambda_{u{\langle i \rangle}}}}
              {{\lambda_x}+w {\lambda_{u{\langle i \rangle}}}}
        \Bigg\}
        +\frac{w {\lambda_{u{\langle i \rangle}}}}{{\lambda_x}+w {\lambda_{u{\langle i \rangle}}}}
        ~{G}_{4,3}^{2,3}
        \left(
        \frac{{\lambda_x} {\lambda_{u{\langle i \rangle}}}}{{\lambda_x}+w {\lambda_{u{\langle i \rangle}}}}
        \left|
        \begin{matrix}
        {1-j-n,1,1,0} \\
        {~~ 1,1,0 }
        \end{matrix}
        \right.
        \right)
        \right]
        \right]
\end{align}
where
$\psi_k \triangleq \frac{1}{2}~ {\psi}\left(k+1\right) +\frac{1}{2} ~{\psi}\left(k+2\right)$,
and
$\mathcal I{_1}\left(w,\lambda\right)
\triangleq
{G}_{3,2}^{1,3}
        \left(
        \frac{{\lambda_x} {\lambda_{u{\langle i \rangle}}}}{{\lambda_x}+w {\lambda_{u{\langle i \rangle}}}}
        \left|
        \begin{matrix}
        {1-j-n,1,1} \\
        {~~ 1,0 }
        \end{matrix}
        \right.
        \right)
        + 2 {\Gamma}\left(j+n\right)
\bigg{\{}
{\ln}\left(\frac{w}{\sqrt{{\lambda_x} {\lambda_y}}}\right)
-\psi_k
\bigg{\}}
$.
    $
    {G}^{m,n}_{p,q}
    \left(
    {x}
    \left|
    \begin{matrix}
    ~{a_p} \\
    ~{b_q}
    \end{matrix}
	\right.
	\right)
    $
    and $\psi\left(x\right)$ are
the Meijer-G function and the Euler psi function defined as \cite[eq.~8.2.1.1]{PBM:90:Book:v3}
and \cite[eq.~8.360.1]{GR:07:Book} respectively.

Note that, exchanging $X$ and $Y$ in \emph{Definition} 1, represents the switching of the interference from relay to destination node in \eqref{IRN:B:01:04}. Hence, the CDF and PDF equations of SM1 are also applicable to the general case when either the relay or destination is subject to interference. In addition, the performance metric will be the same if an equal power sharing is adopted between the source and relay nodes, that is, when $\zeta = 0.5$.

\begin{IEEEproof}
See Appendix~\ref{sec:appendix:A1}.
\end{IEEEproof}

\end{theorem}



\begin{definition}  \label{def:02:SINR:MD02}

Let random variables $X$, $Y$, $U$ and $V$ are statistically independent where $X \sim \ExpRV{\frac{1}{\lambda_x}}$ and $Y \sim \ExpRV{\frac{1}{\lambda_y}}$. Random variables $U=\sum\limits_{l=1}^{L_1}U_l +1$ and $V=\sum\limits_{l=1}^{L_1}V_l +1$, $U_l$ and $V_l$ are i.n.i.d. exponentially distributed random variables with PDFs \cite{BSW:07:WCOM}
\begin{align} \label{PDF:U:MD02}
\PDF{U}{u} =\sum\limits_{i = 1}^{\zeta \left( \Lambda_1 \right)}
            \sum\limits_{j = 1}^{{\tau _i}\left( \Lambda_1 \right)}
            \frac{
            {{X_{i,j}}\left( \Lambda_1 \right)}}
            {\Gamma\left(j\right) \lambda_{u{\langle i \rangle}}^{j}}
            {\left(u-1\right)^{j-1}}
            e^{-\frac{\left(u-1\right)}{\lambda_{u{\langle i \rangle}}}}
            ,~~~~~~~~~~~~~~~~u \geq 1
\end{align}
\begin{align} \label{PDF:V:MD02}
\PDF{V}{v}  =\sum\limits_{p = 1}^{\zeta \left( \Lambda_2 \right)}
            \sum\limits_{q = 1}^{{\tau _i}\left( \Lambda_2 \right)}
            \frac{
            {{X_{p,q}}\left( \Lambda_2 \right)}}
            {\Gamma \left(q\right) \lambda_{v{\langle p \rangle}}^{q}}
            {\left(v-1\right)^{q-1}}
            e^{-\frac{\left(v-1\right)}{\lambda_{v{\langle p \rangle}}}}
            ,~~~~~~~~~~~~~~~~v \geq 1
\end{align}
respectively.
We define a r.v. $Z$ such that
\begin{align}  \label{eq:W:02}
Z=\frac{XY}{XU+YV}
\end{align}

\end{definition}

\begin{theorem}[CDF and PDF] \label{thm02:cdf:pdf}

The cumulative distribution function (CDF) and probability density function (PDF) of random variable $Z$ are given respectively as

\begin{align} \label{IRN:CDF:MD2:01}
&\CDF{Z}{z}
=   1-
    \sum\limits_{i = 1}^{\zeta \left( \Lambda_1 \right)}
    \sum\limits_{j = 1}^{{\tau _i}\left( \Lambda_1 \right)}
    \sum\limits_{p = 1}^{\zeta \left( \Lambda_2 \right)}
    \sum\limits_{q = 1}^{{\tau _p}\left( \Lambda_2 \right)}
    {X_{i,j}}\left( \Lambda_1 \right)
    {X_{p,q}}\left( \Lambda_2 \right)
    e^{-z \left(\frac{1}{\lambda_x}+\frac{1}{{\lambda_y}}\right)}
    \left(\frac{{\lambda_y}}{{\lambda_y}+z {\lambda_{u{\langle i \rangle}}}}\right)^j
\nonumber
\\
&
    \times
    \left(\frac{{\lambda_x}}{{\lambda_x}+z {\lambda_{v{ \langle p \rangle }}}}\right)^q
    \left[
    1+
    \sum \limits_{k=0}^{\infty }
	\sum \limits_{n=0}^{k+1}
	\sum \limits_{r=0}^{k+1}
    \binom{k+1}{n}
    \binom{k+1}{r}
    \frac{z^{2k+2}}
    {\Gamma\left(j\right) \Gamma\left(q\right)
    \left({\lambda_x} {\lambda_y}\right)^{k+1}
    k! \left(k+1\right)!}
\right.
\nonumber
\\
&
\left.
    \times
    \left(\frac{{\lambda_y {\lambda_{u{ \langle i \rangle }}}}}{{\lambda_y}+z {\lambda_{u{ \langle i \rangle }}}}\right)^r
    \left(\frac{{\lambda_x {\lambda_{v{ \langle p \rangle }}}}}{{\lambda_x}+z {\lambda_{v{ \langle p \rangle }}}}\right)^n
    \left[
    2 \Gamma\left(j+r\right)
    \Gamma\left(q+n\right)
    \left\{
        {\ln}\left(\frac{z}{\sqrt{{\lambda_x} {\lambda_y}}}\right)
		-\psi_k
    \right\}
    +\Gamma\left(j+r\right)
\right.
\right.
\nonumber
\\
&
\left.
\left.
        \times
        ~{G}^{1,3}_{3,2}
        \left(
        \frac{{\lambda_x} {\lambda_{v{\langle p \rangle }}} }
        {{\lambda_x}+z {\lambda_{v{\langle p \rangle }}}}
        \left|
        \begin{matrix}
        {1-q-n,1,1} \\
        {~~ 1,0}
        \end{matrix}
        \right.
        \right)
    + \Gamma\left(q+n\right)
        ~{G}^{1,3}_{3,2}
        \left(
        \frac{{\lambda_y} {\lambda_{u{\langle i \rangle }}} }
        {{\lambda_y}+z {\lambda_{u{\langle i \rangle }}}}
        \left|
        \begin{matrix}
        {1-j-r,1,1} \\
        {~~ 1,0}
        \end{matrix}
        \right.
        \right)
    \right]
    \right]
\end{align}
\begin{align} \label{IRN:PDF:MD2:01}
&\PDF{Z}{z}
    =
    \sum\limits_{i = 1}^{\zeta \left( \Lambda_1 \right)}
    \sum\limits_{j = 1}^{{\tau _i}\left( \Lambda_1 \right)}
    \sum\limits_{p = 1}^{\zeta \left( \Lambda_2 \right)}
    \sum\limits_{q = 1}^{{\tau _p}\left( \Lambda_2 \right)}
    {X_{i,j}}\left( \Lambda_1 \right)
    {X_{p,q}}\left( \Lambda_2 \right)
    e^{-z \left(\frac{1}{\lambda_x}+\frac{1}{{\lambda_y}}\right)}
    \left(\frac{{\lambda_y}}{{\lambda_y}+z {\lambda_{u{ \langle i \rangle }}}}\right)^j
    \left(\frac{{\lambda_x}}{{\lambda_x}+z {\lambda_{v{ \langle p \rangle }}}}\right)^q
\nonumber
\\
&
    ~~~~~~~~\times
    \left[
    \left\{
    \frac{1}{\lambda_x}+\frac{1}{{\lambda_y}}
    + \frac{q {\lambda_{v{ \langle p \rangle }}} }{{\lambda_x}+z {\lambda_{v{ \langle p \rangle }}}}
    \!\! +\frac{j {\lambda_{u{ \langle i \rangle }}} }{{\lambda_y}+z {\lambda_{u{ \langle i \rangle }}}}
    \right\}
    -
    \sum \limits_{k=0}^{\infty }
	\sum \limits_{n=0}^{k+1}
	\sum \limits_{r=0}^{k+1}
    \binom{k+1}{n}
    \binom{k+1}{r}
\right.
\nonumber
\\
&
\left.
    ~~~~~~\times
    \frac{z^{2k+1}}
    {\Gamma\left(j\right) \Gamma\left(q\right)
    k! \left(k+1\right)!
    \left({\lambda_x} {\lambda_y}\right)^{k+1}}
    \left(\frac{{\lambda_y {\lambda_{u{ \langle i \rangle }}}}}{{\lambda_y}+z {\lambda_{u{ \langle i \rangle }}}}\right)^r
    \left(\frac{{\lambda_x {\lambda_{v{ \langle p \rangle }}}}}{{\lambda_x}+z {\lambda_{v{ \langle p \rangle }}}}\right)^n
    \Bigg[
    2 \Gamma\left(q+n\right) \Gamma\left(j+r\right)
\right.
\nonumber
\\
&
\left.
\left.
    ~~~~~~~~~~~~~
    +
    \mathcal{I}_2 \left(z,\lambda\right)
    \left\{
    2 k+2 - \frac{z \left(\lambda_x+\lambda_y\right)}{\lambda_x \lambda_y}
    -\frac{\left(j+r\right)z {\lambda_{u{ \langle i \rangle }}}}
        {{\lambda_y}+z {\lambda_{u{ \langle i \rangle }}}}
    -\frac{\left(q+n\right)z {\lambda_{v{ \langle p \rangle }}}}
        {{\lambda_x}+z {\lambda_{v{ \langle p \rangle }}}}
    \right\}
\right.
\right.
\nonumber
\\
&
\left.
\left.
    ~~~~~~~~~~~~~
    +\frac{\Gamma\left(j+r\right) {z \lambda_{v{ \langle p \rangle }}}}
    {{\lambda_x}+z {\lambda_{v{ \langle p \rangle }}}}
        ~{G}^{2,3}_{4,3}
        \left(
        \frac{{\lambda_x} {\lambda_{v{\langle p \rangle }}} }
        {{\lambda_x}+z {\lambda_{v{\langle p \rangle }}}}
        \left|
        \begin{matrix}
        {1-q-n,1,1,0} \\
        {~~ 1,1,0}
        \end{matrix}
        \right.
        \right)
\right.
\right.
\nonumber
\\
&
\left.
\left.
    ~~~~~~~~~~~~~
    +
    \frac{\Gamma\left(q+n\right){z \lambda_{u{ \langle i \rangle }}}}
        {{\lambda_y}+z {\lambda_{u{ \langle i \rangle }}}}
        ~{G}^{2,3}_{4,3}
        \left(
        \frac{{\lambda_y} {\lambda_{u{\langle i \rangle }}} }
        {{\lambda_y}+z {\lambda_{u{\langle i \rangle }}}}
        \left|
        \begin{matrix}
        {1-j-r,1,1,0} \\
        {~~ 1,1,0}
        \end{matrix}
        \right.
        \right)
    \right]
    \right]
\end{align}
where,
\\
$\mathcal{I}_2 \left(z,\lambda\right)
\triangleq 2\Gamma\left(q+n\right) \Gamma\left(j+r\right) \left\{\ln \frac{z}{\sqrt{\lambda_x \lambda_y}}- \psi_k\right\}
        +\Gamma\left(j+r\right)
        ~{G}^{1,3}_{3,2}
        \left(
        \frac{{\lambda_x} {\lambda_{v{\langle p \rangle }}} }
        {{\lambda_x}+z {\lambda_{v{\langle p \rangle }}}}
        \left|
        \begin{matrix}
        {1-q-n,1,1} \\
        {~~ 1,0}
        \end{matrix}
        \right.
        \right)
        \\
        +\Gamma\left(q+n\right)
        ~{G}^{1,3}_{3,2}
        \left(
        \frac{{\lambda_y} {\lambda_{u{\langle i \rangle }}} }
        {{\lambda_y}+z {\lambda_{u{\langle i \rangle }}}}
        \left|
        \begin{matrix}
        {1-j-r,1,1} \\
        {~~ 1,0}
        \end{matrix}
        \right.
        \right)
$

\begin{IEEEproof}
See Appendix~\ref{sec:appendix:B}.
\end{IEEEproof}

\end{theorem}


\begin{proposition} \label{Prop01:cdf:pdf:iidRay}

Consider a system with $L$ interferers in both relay and destination nodes where the interfering channels are i.i.d. Rayleigh faded. Adopting an equal power allocation protocol for source and relay node the cumulative distribution function of ${\gamma_{\mathrm{SINR_{2}}}}$ in \eqref{IRN:B:02:04} can be written as

\begin{align} \label{CDF:MD01:011Ray}
&\CDF{\gamma_{\mathrm{SINR_{2}}}}{\gamma}
=
		1-e^{-{2 \gamma}/\lambda}
		\left(\frac{{\lambda}}{{\lambda}+{\gamma} {\lambda_I}}\right)^{2L}
		-\frac{1}{{\Gamma}\left(L\right)^2}
		~e^{-{2 \gamma}/\lambda}
        \sum \limits_{k=0}^{\infty }
		\sum \limits_{n=0}^{k+1}
        \sum \limits_{r=0}^{k+1}
        \binom{k+1}{n}
        \binom{k+1}{r}
\nonumber
\\
&		\times
		\frac{{\lambda_I}^{n+r}{\gamma}^{2k+2}}
             {{\lambda}^{2k+2} k! (k+1)!}
        \left(\frac{{\lambda}}{{\lambda}+{\gamma} {\lambda_I}}\right)^{2L+n+r}
		\Bigg[
        2 {\Gamma}\left(L+n\right){\Gamma}\left(L+r\right)
		\left\{ \ln \frac{{\gamma}}{\lambda}
		-\psi_k
        \right\}
\nonumber
\\
&
\left.
        +\Gamma\left(L+r\right)
        {G}^{1,3}_{3,2}
        \left(
        \frac{{\lambda} {\lambda_I}}{{\lambda}+{\gamma} {\lambda_I}}
        \left|
        \begin{matrix}
        {1-L-n,1,1} \\
        {~~ 1,0 }
        \end{matrix}
        \right.
        \right)
        +\Gamma\left(L+n\right)
        {G}^{1,3}_{3,2}
        \left(
        \frac{{\lambda} {\lambda_I}}{{\lambda}+{\gamma} {\lambda_I}}
        \left|
        \begin{matrix}
        {1-L-r,1,1} \\
        {~~ 1,0 }
        \end{matrix}
        \right.
        \right)
\right]
\end{align}

\begin{IEEEproof}
See Appendix~\ref{sec:appendix:A11}.
\end{IEEEproof}

\end{proposition}

\section{Interference at Relay: I.I.D. Nakagami-$m$ Interferers}  \label{sec:A:02}

\begin{proposition} \label{thm01:cdf:pdf:Nak}
 Suppose the interfering channels are  i.i.d. Nakagami-$m$ distributed while the main channels (S-R and R-D) are Rayleigh faded. In this case the CDF of ${\gamma_{\mathrm{SINR_{2}}}}$ can be written as

\begin{align} \label{CDF:MD01:01Nak}
&\CDF{\gamma_{\mathrm{SINR_{2}}}}{\gamma}
=   1-
    e^{-{\gamma} \left(\frac{1}{\lambda_1}+\frac{1}{{\lambda_2}}\right)}
    \left(\frac{{\lambda_2}}{{\lambda_2}+{\gamma} {\alpha_{I,1}}}\right)^{{m_1 L_1}}
    \left(\frac{{\lambda_1}}{{\lambda_1}+{\gamma} {\alpha_{I,2}}}\right)^{{m_2 L_2}}
\nonumber
\\
&
    \times
    \left[
    1+
    \sum \limits_{k=0}^{\infty }
	\sum \limits_{n=0}^{k+1}
	\sum \limits_{r=0}^{k+1}
    \binom{k+1}{n}
    \binom{k+1}{r}
    \frac{{\gamma}^{2k+2}
    \left({\lambda_1} {\lambda_2}\right)^{-k-1}}
    {\Gamma\left({m_1 L_1}\right) \Gamma\left({{m_2 L_2}}\right)
    k! \left(k+1\right)!}
    \left(\frac{{\lambda_2 {\alpha_{I,1}}}}{{\lambda_2}+{\gamma} {\alpha_{I,1}}}\right)^r
\right.
\nonumber
\\
&
\left.
    \times
    \left(\frac{{\lambda_1 {\alpha_{I,2}}}}{{\lambda_1}+{\gamma} {\alpha_{I,2}}}\right)^n
    \left[
    2 \Gamma\left({m_1 L_1}+r\right)
    \Gamma\left({{m_2 L_2}}+n\right)
    \left\{
        {\ln}\left(\frac{{\gamma}}{\sqrt{{\lambda_1} {\lambda_2}}}\right)
		-\psi_k
    \right\}
\right.
\right.
\nonumber
\\
&
\left.
\left.
        +\Gamma\left({m_1 L_1}+r\right)
        ~{G}^{1,3}_{3,2}
        \left(
        \frac{{\lambda_1} {\alpha_{I,2}} }
        {{\lambda_1}+{\gamma} {\alpha_{I,2}}}
        \left|
        \begin{matrix}
        {1-{{m_2 L_2}}-n,1,1} \\
        {~~ 1,0}
        \end{matrix}
        \right.
        \right)
\right.
\right.
\nonumber
\\
&
\left.
\left.
    + \Gamma\left({{m_2 L_2}}+n\right)
        ~{G}^{1,3}_{3,2}
        \left(
        \frac{{\lambda_2} {\alpha_{I,1}} }
        {{\lambda_2}+{\gamma} {\alpha_{I,1}}}
        \left|
        \begin{matrix}
        {1-{m_1 L_1}-r,1,1} \\
        {~~ 1,0}
        \end{matrix}
        \right.
        \right)
    \right]
    \right]
\end{align}
where $\alpha_{I,i}=\lambda_{I,i}/m_i$, $i \in \{1,2\}$. $\lambda_{I,1}$ and $\lambda_{I,2}$ are the average INR at the relay and the destination respectively; similarly, $m_1$ and $m_2$ are i.i.d. Nakagami-$m$ parameter at the relay and destination respectively. The average SNR of the $1$st and $2$nd hop are $\lambda_1$ and $\lambda_2$ respectively.

\begin{IEEEproof}
See Appendix~\ref{sec:appendix:A12}.
\end{IEEEproof}

\end{proposition}

\section{Outage Probability}  \label{sec:OP}
In this section the derived analytical results are used to the investigate wireless network outage probability performance. We define outage probability as the probability that the instantaneous receiver SINR falls below a predefined threshold value of SINR $\gamma_{\mathrm{th}}$. We consider $\gamma_{\mathrm{th}}=\rho \left(2^{M R} -1 \right)$, where $\rho$ varies from 1 to 6.4 depending on the degree of coding, $M$ be the number of hops and  $R$ the data rate in bits/s/Hz \cite{GDC:09:TWC}. The CDF equations \eqref{CDF:MD01:01}, \eqref{IRN:CDF:MD2:01} and \eqref{CDF:MD01:01Nak} may be used to evaluate the outage probabilities  in SM 1 and SM 2 with an arbitrary number of interferers and interfering powers.


\section{Numerical Analysis}  \label{sec:D}

In this section, numerical results on the derived equations are presented and compared with Monte-Carlo simulations. Due to lack of transmitter CSI we assume the source and the relay evenly share the total system power $P_\mathrm{tot}$. Furthermore, for fair comparison with relevant studies the average channel gain of all hops in this network are set to unity and the noise variance at the relay and destination nodes are set to $\sigma^2$. We consider $\gamma_{\mathrm{th}}=3$ which achieve a data rate $R=1$ in a dual hop network unless otherwise specified. The maximum value of $k=100$ for sum terms in \eqref{CDF:MD01:01} and \eqref{IRN:CDF:MD2:01} is used, which is sufficient to match the analytical result with the simulation data. The number of interferers at the relay and destination are assumed to be same ($L_1 =L_2$) in SM 2 networks, and all the interferers have equal transmit power. Throughout this section we will use total average transmit SNR, $\lambda_\mathrm{tot} \triangleq P_\mathrm{tot}/\sigma^2$ as a function in performance of SINR plots.

\begin{figure}[!t]
    \centerline{\includegraphics[width=0.8\textwidth]{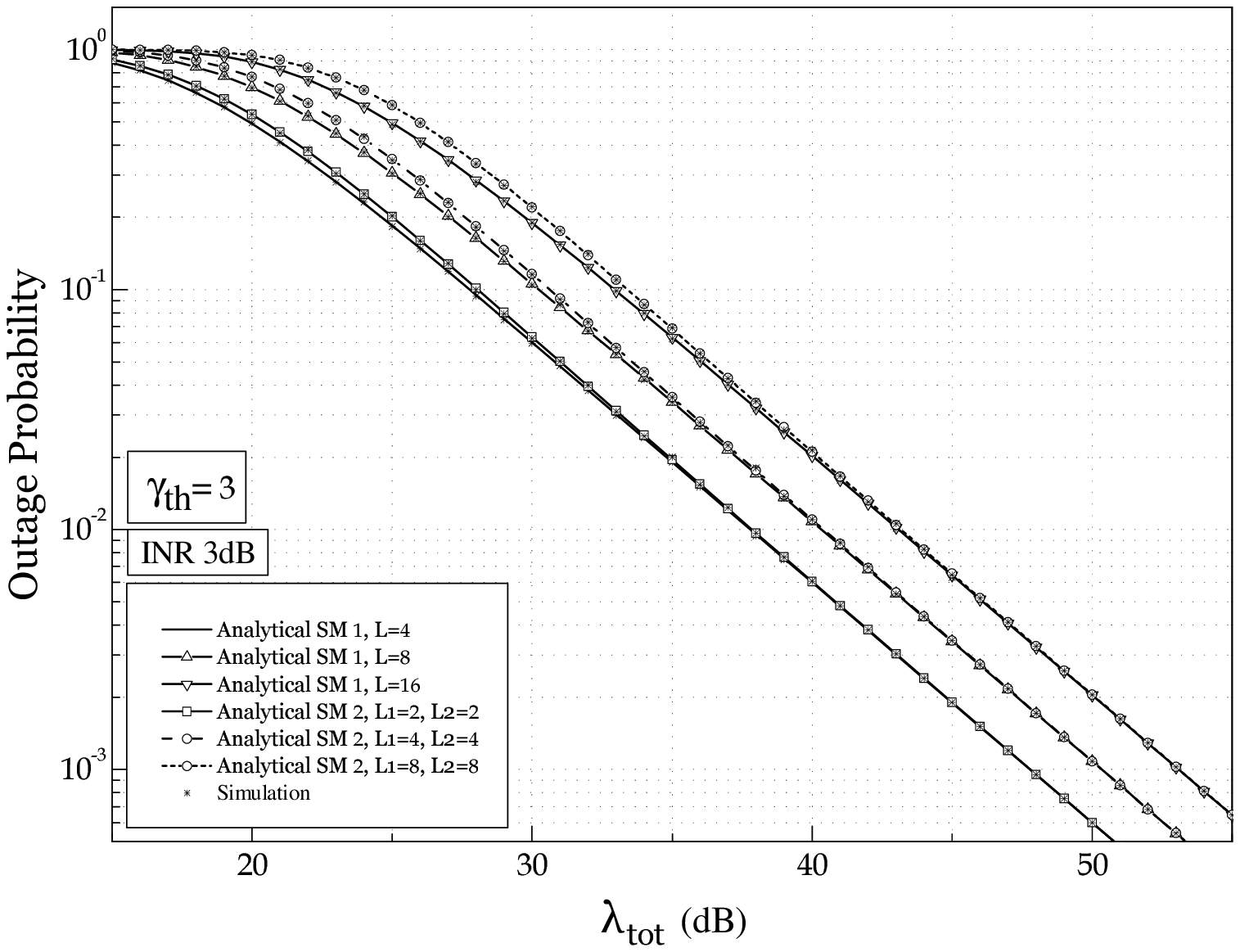}}
    \caption{
    Outage probability vs total SNR in system model 1 and 2.
    }
    \label{fig:B}
\end{figure}

\begin{figure}[!t]
    \centerline{\includegraphics[width=0.75\textwidth]{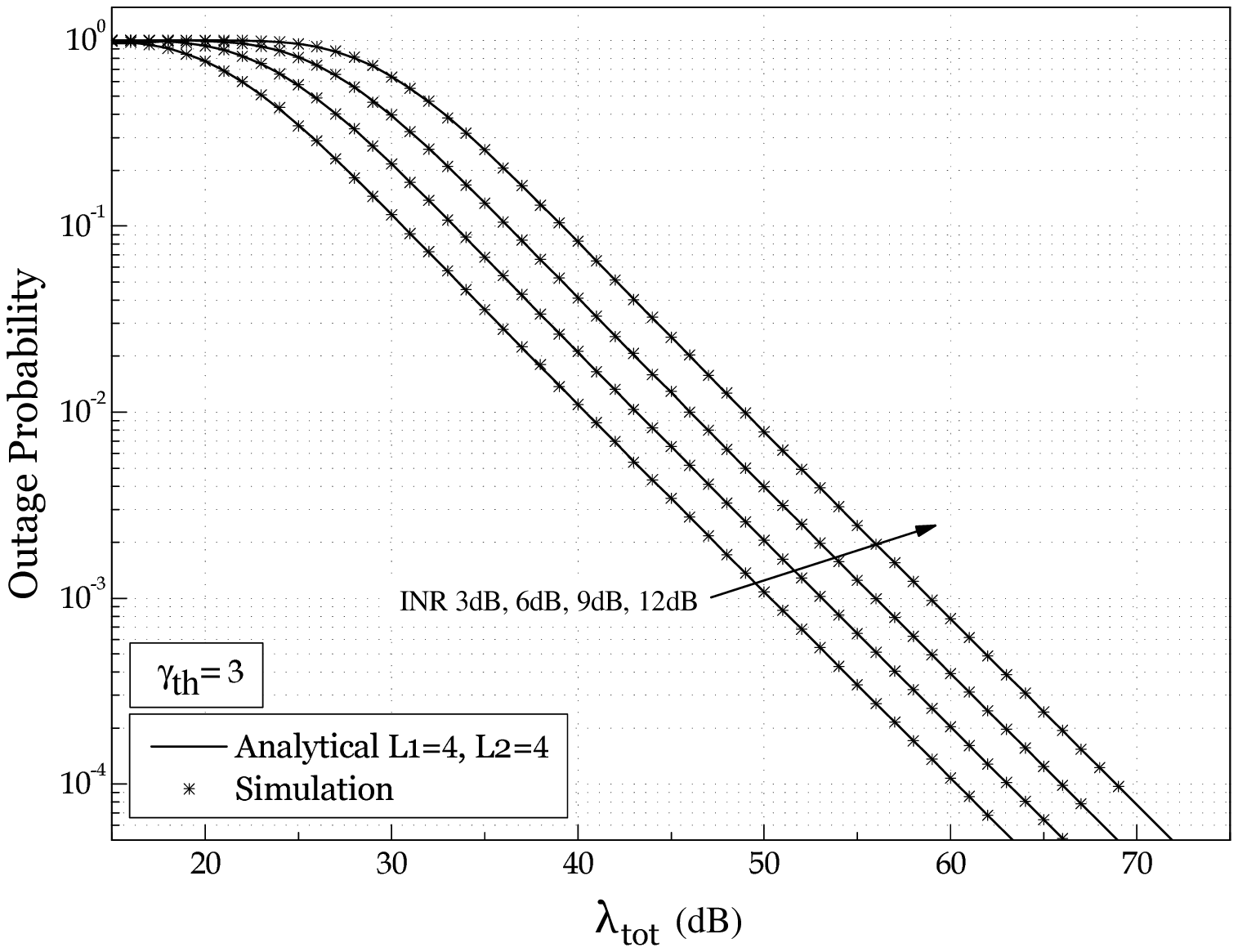}}
    \caption{
    Outage probability vs total SNR in system model 2 with different level of INR's.
    }
    \label{fig:C}
\end{figure}

\begin{figure}[!t]
    \centerline{\includegraphics[width=0.75\textwidth]{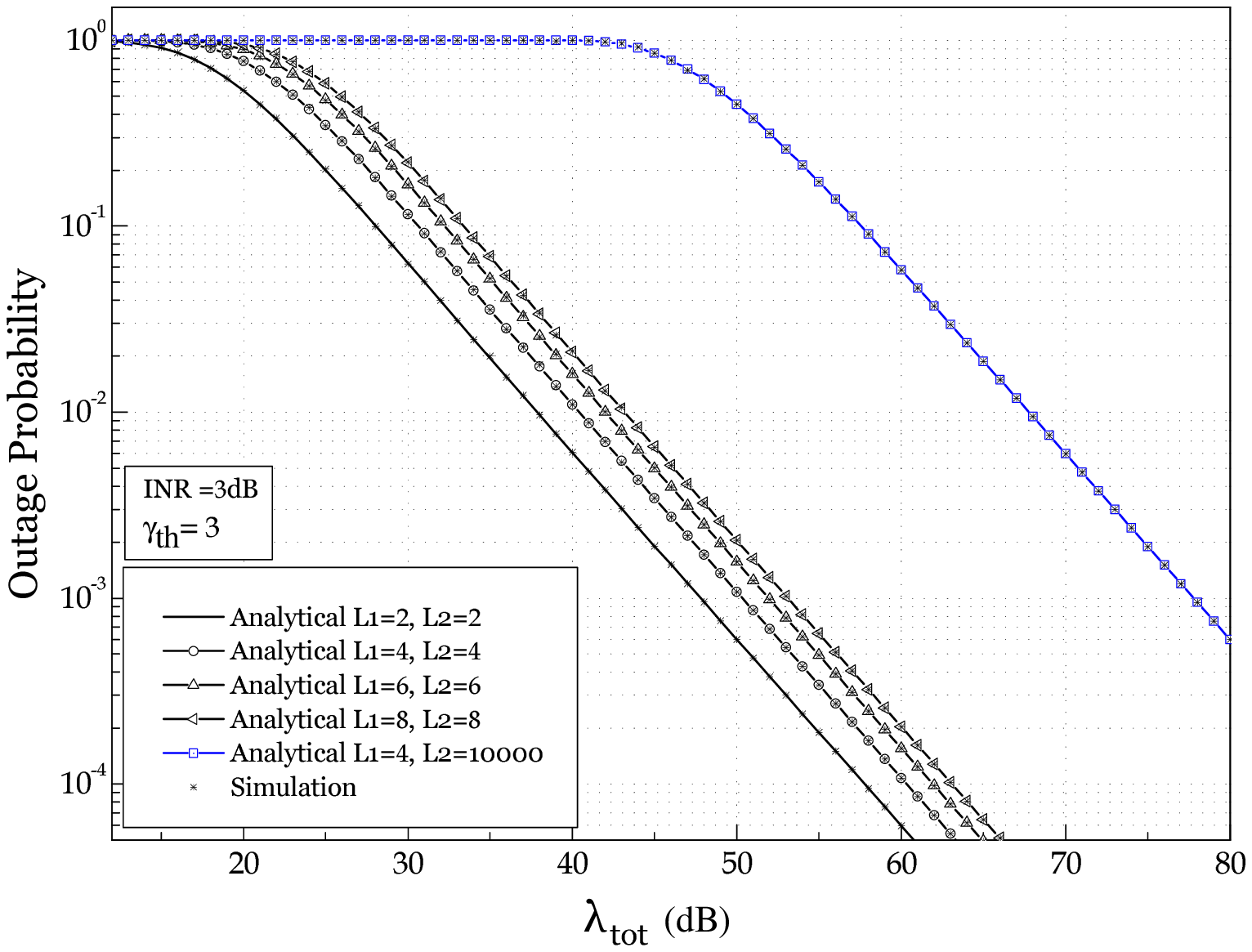}}
    \caption{
    Outage probability vs total SNR in system model 2 with different number of interferers when INR=3 dB.
    }
    \label{fig:D}
\end{figure}

Fig.~\ref{fig:B} shows the outage probability (OP) as a function of total transmit SNR for a total of $\{4, 8, 16\}$ interferers in SM 1 and SM 2 with INR fixed at 3 dB. All wireless channels (main channels and the interfering channels) are assumed to be Rayleigh faded. As expected, outage probability increases with an increase in the number of interferers. However, when SM 1 has the same number of interferers as the total number of interferers for SM 2, SM1 shows slightly better performance compared to SM2 at a moderately low SINR region only. To study the OP performance with different interfering powers Fig. \ref{fig:C} shows outage probability as a function of $\lambda_\mathrm{tot}$ for a system with 4 interferers at the relay and 4 interferers at the destination. The INR values \{3, 6, 9, 12\} dB are considered. Interestingly, the figure suggests that every 3 dB increase in average INR values requires a subsequent increase of 3 dB total SNR on average to ensure an OP of around $10^{-3}$. In contrast, Fig.~\ref{fig:D} shows OP in a fixed 3 dB INR system in SM 2 with a total of \{4, 8, 12, 16\} number of interferers at the relay and destination respectively. To see how the interference system behaves if one of the nodes experiences most of the interference, Fig.~\ref{fig:D} shows OP with a large $L_2$ and $L_1=4$. For very high interference levels e.g. $L_2=10000$ interferers at the destination node, a $\lambda_\mathrm{tot}$ = 77 dB is required to achieve an outage probability of $10^{-3}$ compared to 50dB when $L_2=4$.

\begin{figure}[!t]
    \centerline{\includegraphics[width=0.7\textwidth]{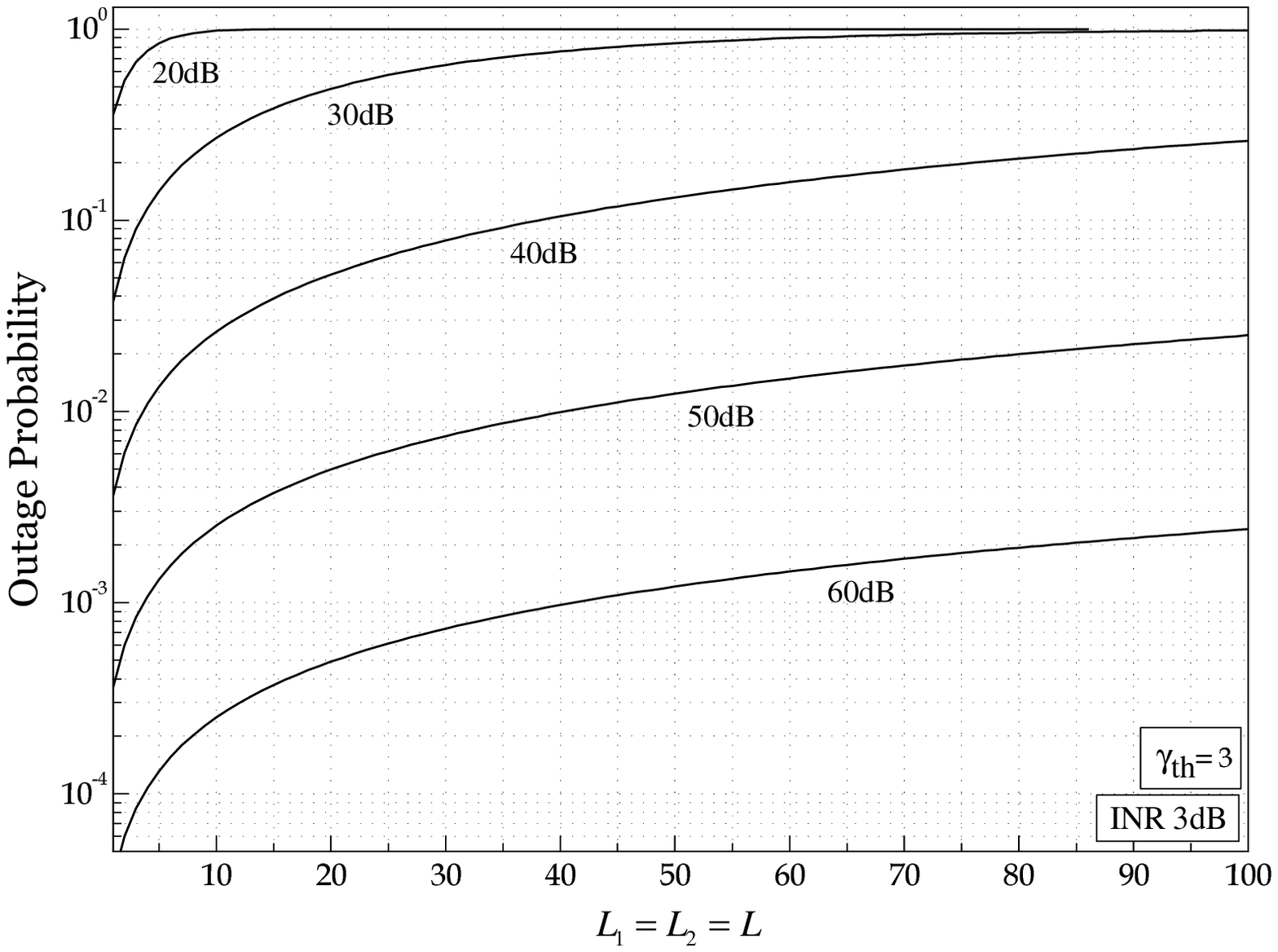}}
    \caption{
    Outage probability as a function of total interferers at the relay and destination when the INR of each interferer is 3 dB and total transmit SNR varies from 20 to 60 dB.
    }
    \label{fig:H}
\end{figure}

Fig.~\ref{fig:H} shows outage probability with number of interferers $L$ in relay and destinations where $L_1=L_2=L$ and each interferer has an INR of 3dB. The figure shows how the outage probability increases exponentially as the number of interferers increases. The plot includes graphs for transmit SNRs of 20, 30, 40, 50 and 60 dB. A very interesting phenomenon observed is that at 60dB of total transmit SNR the system can support up to 40 interferers at the relay and destination, ensuring an outage probability of $10^{-3}$, however, with a reduction of only 10dB in total transmit SNR the performance reduces drastically and it can support only 4 interferers at the relay and destination for the same outage probability.

\begin{figure}[!t]
    \centerline{\includegraphics[width=0.75\textwidth]{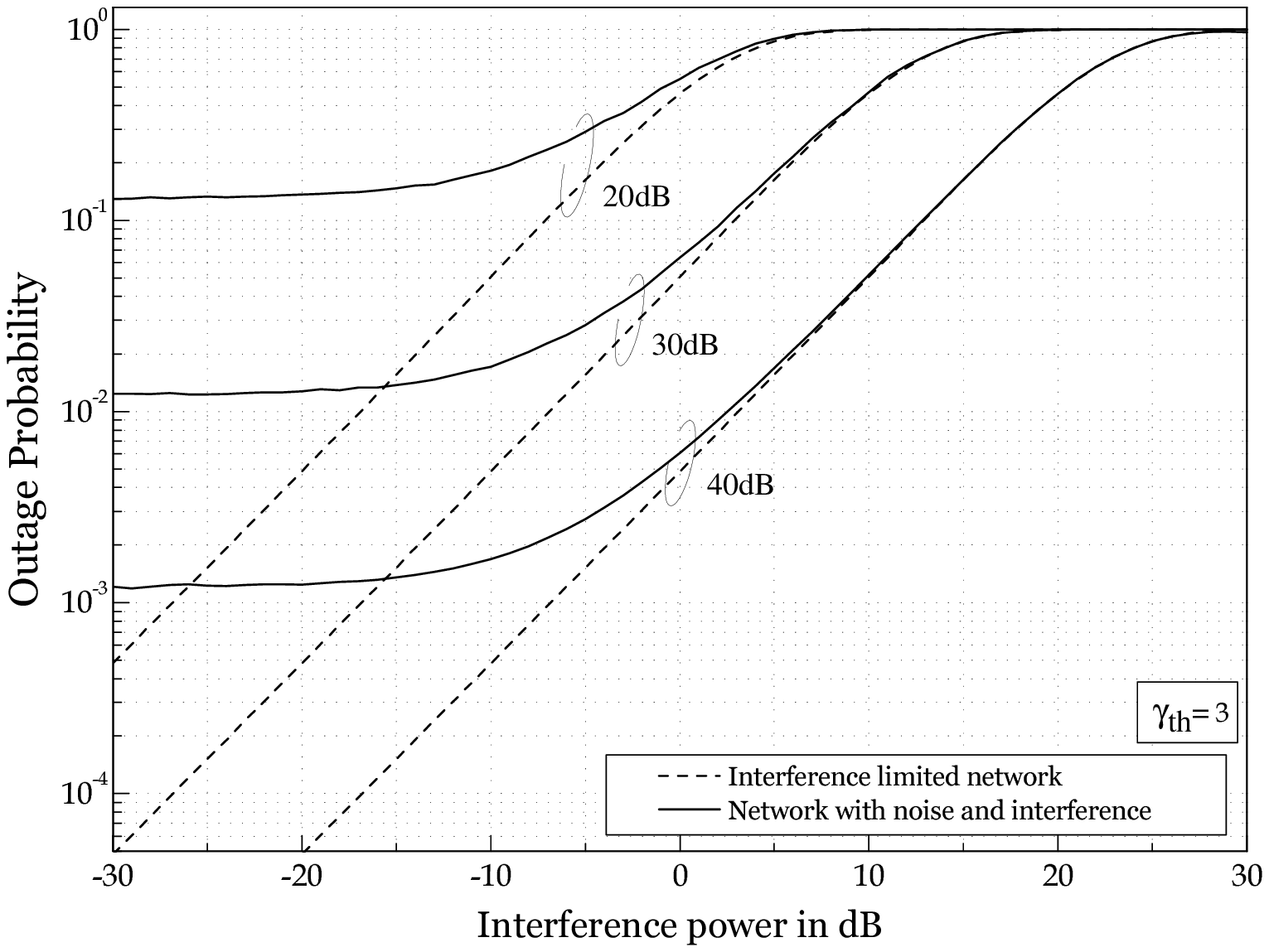}}
    \caption{
    Outage probability as a function of interference power when the relay and the destination have $4$ interferers and $\lambda_{\mathrm{tot}}$ is 20, 30 and 40 dB's.
    }
    \label{fig:I}
\end{figure}

Fig.~\ref{fig:I} shows the outage probability as a function of interference power when the relay and destination are both subject to $4$ interferers. The figure compares the outage probability of an interference limited system (dashed line) studied in \cite{LL:11:TVT} to a system where both noise and interference are present. To express total transmit power and interference power in dB we normalize the terms considering noise power equal to 1 unit. Total transmission power of 20, 30 and 40 dB  are considered in the figure and the power of each interferer is rated from -30dB to +30dB. Figure shows that the outage probability the noise plus interference network will be similar to an interference limited network when the transmission power of each interferer is more than 10 dB. However, performance will vary if the number of interferers in the network changes. The figure also illustrates how noise dominates outage probability metric when interference power is low.

\begin{figure}[!t]
    \centerline{\includegraphics[width=0.8\textwidth]{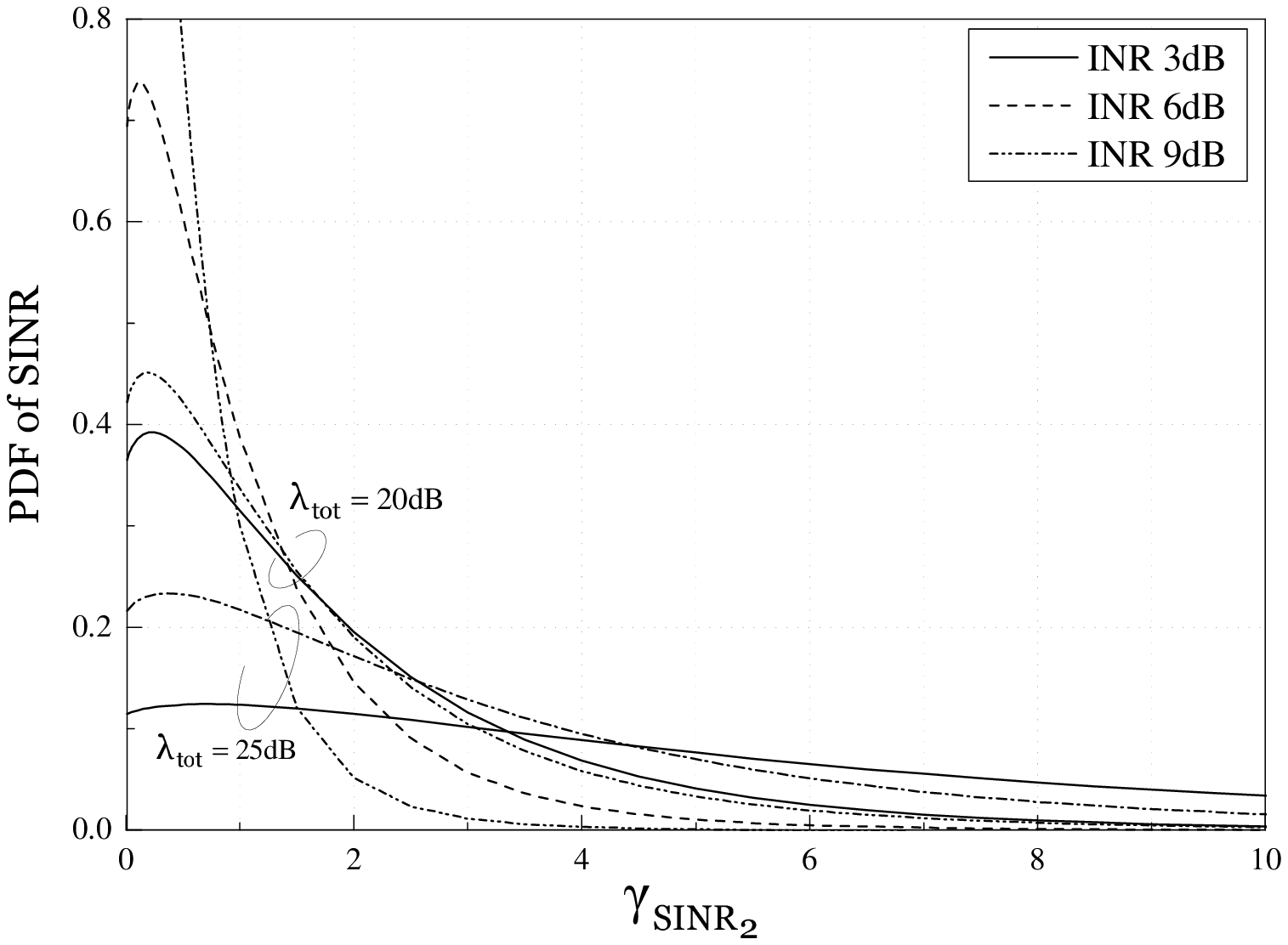}}
    \caption{
    Probability density function of SINR of system model 2.
    }
    \label{fig:E}
\end{figure}

\begin{figure}[!t]
    \centerline{\includegraphics[width=0.8\textwidth]{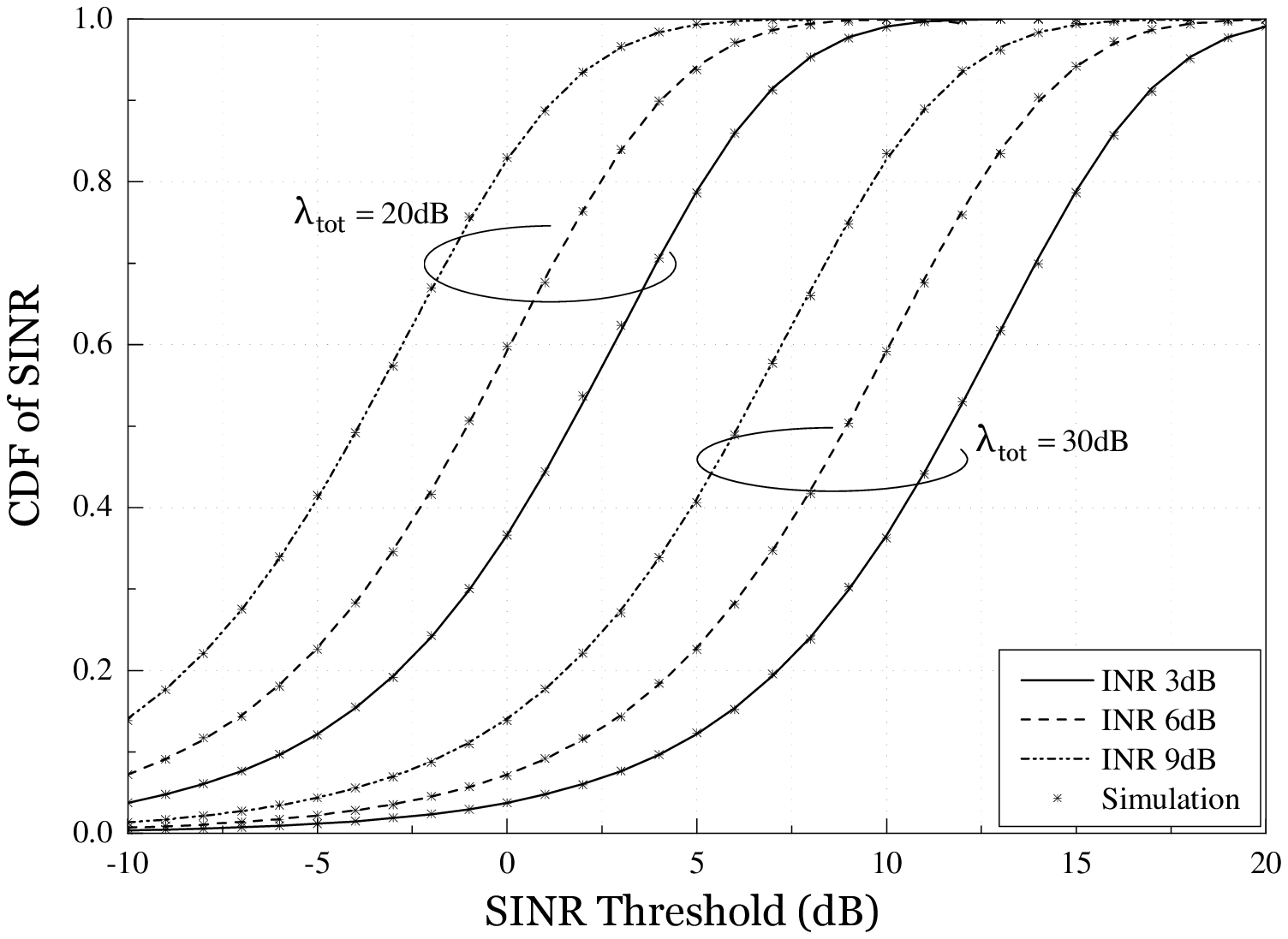}}
    \caption{
    Cumulative distribution function of SINR of system model 2.
    }
    \label{fig:F}
\end{figure}

The PDF of $\gamma_\mathrm{SINR_2}$ is presented in Fig.~\ref{fig:E} for two different $\lambda_\mathrm{tot}$, 20 dB and 25 dB, when the individual interferer INRs are \{3, 6, 9\} dB. The figure implies that a lower INR per interferer increases the probability of higher output SINR at the receiver and similarly when the INR is constant, an increase in average total signal power results in a higher output SINR. In addition, Fig.~\ref{fig:F} shows the CDF of $\gamma_\mathrm{SINR_2}$ as a function of $\gamma_\mathrm{th}$ in dB for two $\lambda_\mathrm{tot}$ groups, 20 dB and 30 dB. Each group contains plots for 3 INRs 3, 6 and 9 dB. It reveals that higher average total SNR and lower INR per interferer decreases the probability of outage.

\begin{figure}[!t]
    \centerline{\includegraphics[width=0.8\textwidth]{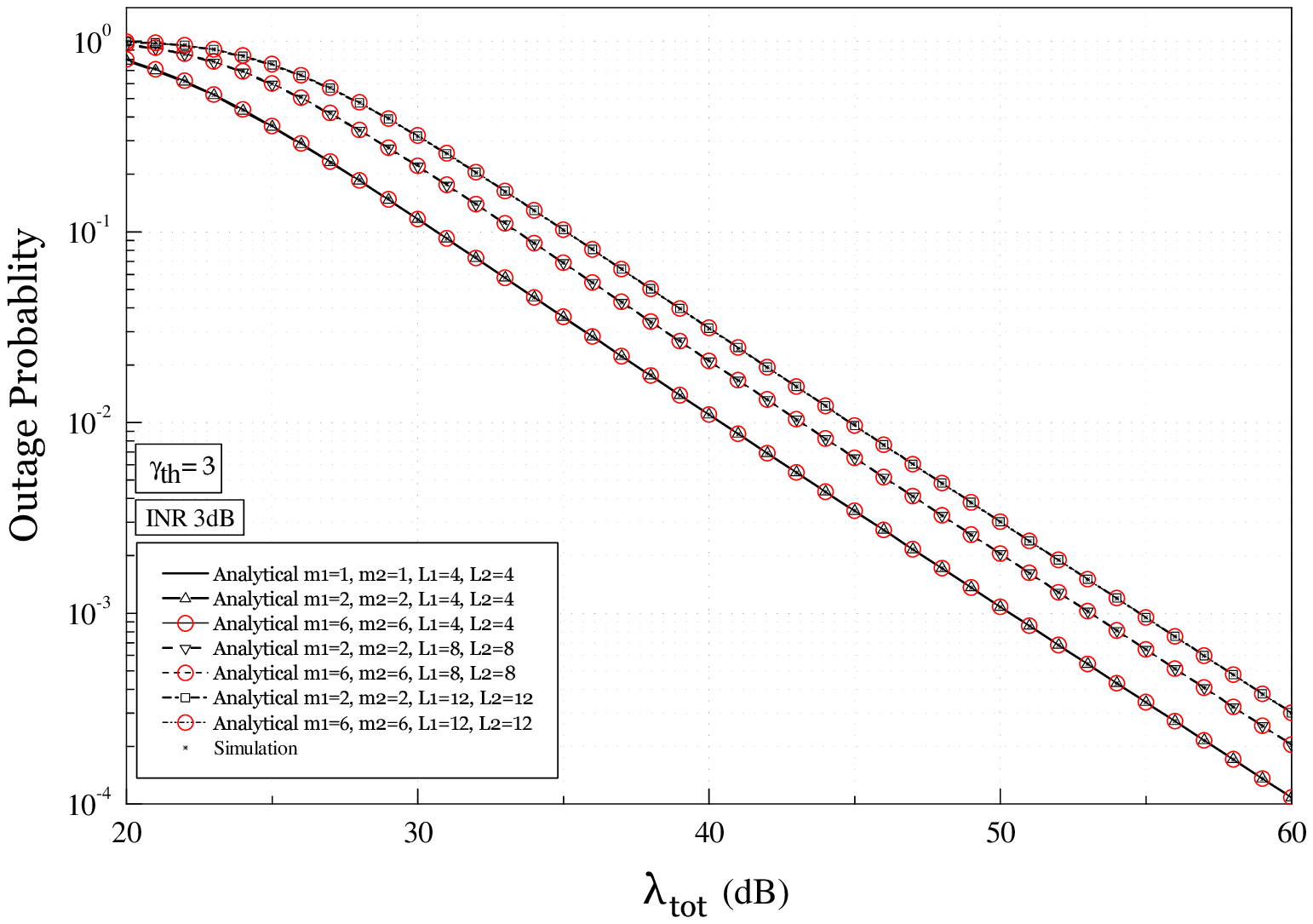}}
    \caption{
    Outage probability vs total SNR over Nakagami-$m$ faded interfering channels in system model 2 with different number of interferers.
    }
    \label{fig:G}
\end{figure}

Finally, Fig.~\ref{fig:G} presents a plot over Nakagami-$m$ faded interfering channels. The outage probability is plotted as a function of $\lambda_{\mathrm{tot}}$ for different number of interferers and different Nakagami channel parameter $m$ in SM 2. Surprisingly, varying the Nakagami $m$ parameter does not result in any significant effects on outage performance if the number of interferers and INR remain constant in the network.


\section{Conclusion}  \label{sec:E}
In this work, we derive the cumulative distribution function and the probability density function of a dual hop interference relay network with arbitrary number of interferers, where the main channels and the interfering channels are i.n.i.d. Rayleigh faded. Later the analysis is extended for i.i.d. Nakagami-$m$ faded interfering channels. This analysis has been performed for two system models; in SM 1 interference is only at the relay node while in SM 2, interference is at both the relay and destination nodes. The derived CDF and PDF expressions for SM 1 can also be used for analyzing an interference network where interference occurs only at the destination node. Numerical results for outage probability performance for different network parameter configurations are presented. The Monte-Carlo simulations show an exact match with the analytical expressions. Results indicate that in a total INR constrained interference network, the number of interferers do not affect the performance of the system in a large scale. Similarly, if the interfering channels are Nakagami-$m$ distributed, changing in Nakagami-parameter $m$ does not affect the performance of the system if the number of interferers and the INR per interferer remain constant.


\appendix

\subsection{Proof of Theorem~\ref{thm01:cdf:pdf:SINR1}}
\label{sec:appendix:A1}

According to the PDF of $X$, $Y$ and $U$ as defined in \emph{Definition} 1, the random variables $X$, $Y$ and $U$ are nonnegative, thus $\CDF{W}{w}=0$ for $w<0$. For $w \geq 0$  we have
\begin{align} \label{eq:CDF:W:Proof:01}
    &\CDF{W}{w}
    =
        \Prob{
            \frac{X Y}{X+YU}
            \leq w
        }
    \nonumber \\
    &~
    =
        \Prob{
            X \left(Y-w\right)
            \leq
            wYU
        }
    \nonumber \\
    &~
    =
    1-e^{-w/\lambda_{y}}
    +
    \int_{1}^{\infty}
    \int_{w}^{\infty}
    \int_{0}^{\frac{wyu}{y-w}}
    \PDF{X}{x}
    \PDF{Y}{y}
    \PDF{U}{u}
    dx dy du
\end{align}

The two inner integrals in \eqref{eq:CDF:W:Proof:01} can easily be solved by using \cite[eq.~3.351.3]{GR:07:Book} and \cite[eq.~3.471.9]{GR:07:Book} and after some manipulation the CDF of $W$ is obtained,

\begin{align} \label{eq:CDF:W:Proof:02}
\CDF{W}{w}
    &=
    1-
    \sum\limits_{i = 1}^{\zeta \left( \Lambda \right)}
    \sum\limits_{j = 1}^{{\tau _i}\left( \Lambda \right)}
    \frac{2 w
    {X_{i,j}}\left( \Lambda \right)
    e^{-w/\lambda_y}
    }
    {\Gamma \left( {j} \right)
    \lambda_{u \langle i \rangle}^j
    \sqrt{\lambda_x \lambda_y}
    }
    \int_{1}^{\infty}
    \sqrt{u}
    {{\left(u-1\right)}^{j - 1}}
    e^{-{{\left(u-1\right)}/ \lambda_{u \langle i \rangle }}}
\nonumber
\\
& ~~~~~~~~~~~~~~~~~~~~~~
    \times
    e^{-w u/\lambda_x}
    K_1 \left(2w \sqrt{\frac{u}{\lambda_x \lambda_y}} \right)
    du
\end{align}
where $K_\nu\left(x\right)$ is the $\nu$th order modified Bessel
function of the second kind. Using \cite[eq.~8.446]{GR:07:Book} we expand the first order modified Bessel function of second kind, then \cite[eq.~8.4.6.5]{PBM:90:Book:v3}, \cite[eq.~3.40.1.1]{PBM:92:Book:v4} and \cite[eq.~3.351.3]{GR:07:Book} are applied to solve the related integrals. After some manipulation we arrive at the  desired result \eqref{CDF:MD01:01}. The PDF of $W$ follows directly from the differentiation of the CDF of $W$ in eq. \eqref{CDF:MD01:01} w.r.t. $w$ by using \cite[eq.~8.2.2.30]{PBM:90:Book:v3},
\begin{align} \label{8.2.2.30PRu}
    \frac{d }{dx}~
    \left[x^\sigma {G}^{m,n}_{p,q}
    \left( {x}
    \left|
    \begin{matrix}
    ~{a_p} \\
    ~{b_q}
    \end{matrix}
	\right.
	\right)
	\right]
=
    -x^{\sigma -1}
    {G}^{m+1,n}_{p+1,q+1}
    \left(
    {x}
    \left|
    \begin{matrix}
    ~{a_p, -\sigma}\\
    ~{1-\sigma, b_q}
    \end{matrix}
	\right.
	\right)
\end{align}
%



\subsection{Proof of Theorem~\ref{thm02:cdf:pdf}}
\label{sec:appendix:B}

Following a similar argument as used in the proof of \emph{Theorem} \ref{thm01:cdf:pdf:SINR1}, the random variables $X$, $Y$, $U$ and $V$ are nonnegative, and thus $\CDF{Z}{z}=0$ for $z<0$. For $z \geq 0$,
\begin{align} \label{eq:CDF:W:Proof:03}
    &\CDF{Z}{z}
    =
        \Prob{
            \frac{X Y}{XU+YV}
            \leq z
        }
    \nonumber \\
    &~
    =
        \Prob{
            X \left(Y-zU\right)
            \leq
            zYV
        }
\\
\nonumber
    &~
    =
\underbrace{    \int_{1}^{\infty}
    \int_{0}^{uz}
    \PDF{Y}{y}
    \PDF{U}{u}
    dy du}
    _{\triangleq~ I_{uy}}
    +
\underbrace{    \int_{1}^{\infty}
    \int_{1}^{\infty}
    \int_{uz}^{\infty}
    \int_{0}^{\frac{zyv}{y-zu}}
    \PDF{X}{x}
    \PDF{Y}{y}
    \PDF{V}{v}
    \PDF{U}{u}
    dx dy dv du}
    _{\triangleq~ I_{uvyx}}
\end{align}
The first part of this expression, integral $I_{uy}$ can be solved by using \cite[eq.~3.351.3]{GR:07:Book}, resulting in

\begin{align} \label{eq:CDF:W:Proof:02:1}
I_{uy}
    =
    1-
    \sum\limits_{i = 1}^{\zeta \left( \Lambda_1 \right)}
    \sum\limits_{j = 1}^{{\tau _i}\left( \Lambda_1 \right)}
    {X_{i,j}}\left( \Lambda_1 \right)
    e^{-z/\lambda_y}
    \left( \frac{\lambda_y}
                {\lambda_y+z\lambda_{u \langle i \rangle}}
    \right)^j
\end{align}
In $I_{uvyx}$, the first two integrals for variables $x$ and $y$ can be solved using a similar approach, via \cite[eq.~3.351.3]{GR:07:Book}, \cite[eq.~3.471.9]{GR:07:Book}, \cite[eq.~8.446]{GR:07:Book}, \cite[eq.~8.4.6.5]{PBM:90:Book:v3} and \cite[eq.~3.40.1.1]{PBM:92:Book:v4}. Substituting \eqref{eq:CDF:W:Proof:02:1} in \eqref{eq:CDF:W:Proof:03} and after some manipulation we have the CDF of $Z$
\begin{align} \label{eq:CDF:W:Proof:02:2}
&\CDF{Z}{z}
    =1
    -
    \sum\limits_{i = 1}^{\zeta \left( \Lambda_1 \right)}
    \sum\limits_{j = 1}^{{\tau _i}\left( \Lambda_1 \right)}
    \sum\limits_{p = 1}^{\zeta \left( \Lambda_2 \right)}
    \sum\limits_{q = 1}^{{\tau _p}\left( \Lambda_2 \right)}
    \frac{
        {X_{i,j}}\left( \Lambda_1 \right)
        {X_{p,q}}\left( \Lambda_2 \right)
        }
        {\Gamma\left(j\right) \Gamma\left(q\right)
        \lambda_{u{\langle i \rangle}}^{j}
        \lambda_{v{\langle p \rangle}}^{q}
        }
    \left[
    e^{-z/\lambda_x} \Gamma\left(q\right)
    \left(\frac{{\lambda_x {\lambda_{v{ \langle p \rangle }}}}}
    {{\lambda_x}+z {\lambda_{v{\langle p \rangle}}}}\right)^q
\right.
\nonumber
\\
&
\left.
    \times
    \int_{1}^{\infty}
    \!\!\!
    \left(u-1\right)^{j-1}
    e^{-\frac{u-1}{{\lambda_{u{\langle i \rangle}}}}}
    e^{-uz/\lambda_y}
    du
    +
    \sum \limits_{k=0}^{\infty }
	\sum \limits_{n=0}^{k+1}
    \binom{k+1}{n}
    \frac{z^{2k+2} e^{-z/\lambda_x}}
        {k! \left(k+1\right)! \left(\lambda_x\lambda_y\right)^{k+1}}
    \left(\frac{{\lambda_x {\lambda_{v{ \langle p \rangle }}}}}
    {{\lambda_x}+z {\lambda_{v{\langle p \rangle}}}}\right)^{q+n}
\right.
\nonumber
\\
&
\left.
    \times
    {G}^{1,3}_{3,2}
    \left(
    \frac{{\lambda_x} {\lambda_{v{\langle p \rangle }}} }
        {{\lambda_x}+z {\lambda_{v{\langle p \rangle }}}}
    \left|
    \begin{matrix}
    {1-q-n,1,1} \\
    {~~ 1,0}
    \end{matrix}
    \right.
    \right)
    \int_{1}^{\infty}
    u^{k+1}
    \left(u-1\right)^{j-1}
    e^{-\frac{u-1}{{\lambda_{u{\langle i \rangle}}}}}
    e^{-uz/\lambda_y}
    du
\right.
\nonumber
\\
&
~~~~\left.
    +
    \sum \limits_{k=0}^{\infty }
	\sum \limits_{n=0}^{k+1}
    \binom{k+1}{n}
    \frac{2 z^{2k+2} e^{-z/\lambda_x}
        \Gamma\left(q+n\right)}
        {k! \left(k+1\right)! \left(\lambda_x\lambda_y\right)^{k+1}}
    \left(\frac{{\lambda_x {\lambda_{v{ \langle p \rangle }}}}}
    {{\lambda_x}+z {\lambda_{v{\langle p \rangle}}}}\right)^{q+n}
\right.
\nonumber
\\
&
~~~~\left.
    \times
    \int_{1}^{\infty}
    u^{k+1}
    \left(u-1\right)^{j-1}
    e^{-\frac{u-1}{{\lambda_{u{\langle i \rangle}}}}}
    e^{-uz/\lambda_y}
    \left\{
    \ln{\sqrt{\frac{u}{\lambda_x \lambda_y}}z}
    -\psi_k
    \right\}
    du
    \right]
\end{align}
where
$\psi_k \triangleq \frac{1}{2}~ {\psi}\left(k+1\right) +\frac{1}{2} ~{\psi}\left(k+2\right)$.
Now using \cite[eq.~3.351.3]{GR:07:Book}, \cite[eq.~8.4.6.5]{PBM:90:Book:v3} and \cite[eq.~3.40.1.1]{PBM:92:Book:v4} it is possible to solve \eqref{eq:CDF:W:Proof:02:2}, which then results in the desired eq. \eqref{IRN:CDF:MD2:01}. Eq \eqref{IRN:PDF:MD2:01} directly follows the differentiation of eq.\eqref{IRN:CDF:MD2:01} utilizing the property \cite[eq.~8.2.2.30]{PBM:90:Book:v3}.


\subsection{Proof of Proposition~\ref{Prop01:cdf:pdf:iidRay}}
\label{sec:appendix:A11}
When the interfering signals are from $L$ equal power sources with i.i.d. interfering channels, $\lambda_{u\langle i \rangle} = \lambda_u = \lambda_I$, $\lambda_{v\langle p \rangle} = \lambda_v = \lambda_I$ and the characteristic coefficient ${{X_{i,j}}\left( \Lambda \right)}$ of $\Lambda$ becomes \cite{BSW:07:WCOM},
          \begin{align}
          {{X_{i,j}}\left( \Lambda \right)}
          =
          \left\{
          \begin{array}{cl}
            0 &~~~ j=1,2,3.. \\
            1 &~~~ j=L
          \end{array}
          \right.
          \end{align}
Using the above facts the proof of the \emph{Proposition} \ref{Prop01:cdf:pdf:iidRay} immediately follows from \emph{Theorem} \ref{thm02:cdf:pdf}. Due to equal power allocation, average 1st and 2nd hop SNR $\lambda_1 = \lambda_2 =\lambda$.


\subsection{Proof of Proposition~\ref{thm01:cdf:pdf:Nak}}
\label{sec:appendix:A12}
If the interfering channels are i.i.d. Nakagami-$m$ faded, in \emph{Definition} \ref{def:02:SINR:MD02}, the PDF of r.v. $U$ and $V$ can be modified to,
\begin{align} \label{PDF:U:MD02:Nak}
\PDF{U}{u}
            =
            \frac
            {\left(u-1\right)^{L_1 m_1-1}}
            {\Gamma\left(L_1 m_1\right) \alpha_{u}^{L_1 m_1}}
            e^{-\frac{\left(u-1\right)}{\alpha_{u}}}
            ,~~~~~~~~~~~~~~~~u \geq 1
\end{align}
and
\begin{align} \label{PDF:V:MD02:Nak}
\PDF{V}{v}
            =
            \frac
            {\left(v-1\right)^{L_2 m_2-1}}
            {\Gamma\left(L_2 m_2\right) \alpha_{v}^{L_2 m_2}}
            e^{-\frac{\left(v-1\right)}{\alpha_{v}}}
            ,~~~~~~~~~~~~~~~~v \geq 1
\end{align}
respectively, where $\alpha_{u}=\lambda_u/m_u$ and $\alpha_{v}=\lambda_v/m_v$ respectively, and $m$ is the corresponding Nakagami-$m$ channel parameter. The r.v.s $X$ and $Y$ are exponentially distributed with hazard rate $1/{\lambda_{x}}$ and $1/{\lambda_{y}}$ respectively. The CDF of $\CDF{\gamma_{\mathrm{SINR_{2}}}}{\gamma}$ for i.i.d. Nakagami-$m$ interfering channels can be obtained by replacing $\lambda_u = \alpha_u$, $\lambda_v = \alpha_v$, $j=L_1 m_1$ and $q=L_2 m_2$ in \eqref{IRN:CDF:MD2:01}.


\bibliographystyle{IEEEtran}
\bibliography{baruabiblio}

\begin{thebibliography}{10}
\providecommand{\url}[1]{#1}
\csname url@samestyle\endcsname
\providecommand{\newblock}{\relax}
\providecommand{\bibinfo}[2]{#2}
\providecommand{\BIBentrySTDinterwordspacing}{\spaceskip=0pt\relax}
\providecommand{\BIBentryALTinterwordstretchfactor}{4}
\providecommand{\BIBentryALTinterwordspacing}{\spaceskip=\fontdimen2\font plus
\BIBentryALTinterwordstretchfactor\fontdimen3\font minus
  \fontdimen4\font\relax}
\providecommand{\BIBforeignlanguage}[2]{{%
\expandafter\ifx\csname l@#1\endcsname\relax
\typeout{** WARNING: IEEEtran.bst: No hyphenation pattern has been}%
\typeout{** loaded for the language `#1'. Using the pattern for}%
\typeout{** the default language instead.}%
\else
\language=\csname l@#1\endcsname
\fi
#2}}
\providecommand{\BIBdecl}{\relax}
\BIBdecl

\bibitem{Meu:71:AAP}
E.~C. van~der Meulen, ``Three-terminal communication channels,'' \emph{Advances
  in Applied Probability}, vol.~3, no.~1, pp. 120--154, 1971.

\bibitem{LTW:04:IT}
J.~N. Laneman, D.~N.~C. Tse, and G.~W. Wornell, ``Cooperative diversity in
  wireless networks: Efficient protocols and outage behavior,'' \emph{IEEE
  Trans. on Inf. theory}, vol.~50, no.~12, pp. 3062--3080, Dec. 2004.

\bibitem{HA:03:WCOM}
M.~O. Hasna and M.-S. Alouini, ``End-to-end performance of transmission systems
  with relays over rayleigh-fading channels,'' \emph{IEEE Transactions on
  Wireless Communications}, vol.~2, no.~6, pp. 1126--1131, Nov. 2003.

\bibitem{HA:04:COM}
------, ``Harmonic mean and end-to-end performance of transmission systems with
  relays,'' \emph{IEEE Transactions on Communications}, vol.~52, no.~1, pp.
  130--135, Jan. 2004.

\bibitem{BHS:08:CL}
B.~Barua, H.~Ngo, and H.~Shin, ``On the sep of cooperative diversity with
  opportunistic relaying,'' \emph{IEEE Communications Letters}, vol.~12,
  no.~10, pp. 727--729, Oct. 2008.

\bibitem{AC:75:IT}
A.~B. Carleial, ``A case where interference does not reduce capacity,''
  \emph{IEEE Transactions on Information Theory}, vol.~21, no.~5, pp. 569--570,
  sep 1975.

\bibitem{AC:78:IT}
------, ``Interference channels,'' \emph{IEEE Transactions on Information
  Theory}, vol.~24, no.~1, pp. 60--70, jan 1978.

\bibitem{RB:79:IT}
R.~Benzel, ``The capacity region of a class of discrete additive degraded
  interference channels,'' \emph{IEEE Transactions on Information Theory},
  vol.~25, no.~2, pp. 228--231, mar 1979.

\bibitem{HK:81:IT}
T.~Han and K.~Kobayashi, ``A new achievable rate region for the interference
  channel,'' \emph{IEEE Transactions on Information Theory}, vol.~27, no.~1,
  pp. 49--60, jan 1981.

\bibitem{MC:85:IT}
M.~Costa, ``On the gaussian interference channel,'' \emph{IEEE Transactions on
  Information Theory}, vol.~31, no.~5, pp. 607--615, sep 1985.

\bibitem{AV:11:IT}
V.~S. Annapureddy and V.~V. Veeravalli, ``Sum capacity of mimo interference
  channels in the low interference regime,'' \emph{IEEE Transactions on
  Information Theory}, vol.~57, no.~5, pp. 2565 --2581, may 2011.

\bibitem{SGN:10:CL}
H.~A. Suraweera, H.~K. Garg, and A.~Nallanathan, ``Performance analysis of two
  hop amplify-and-forward systems with interference at the relay,'' \emph{IEEE
  Communications Letters}, vol.~14, no.~8, pp. 692 --694, Aug. 2010.

\bibitem{QDZQA:11:SPL}
F.~S. Al-Qahtani, T.~Q. Duong, C.~Zhong, K.~A. Qaraqe, and H.~Alnuweiri,
  ``Performance analysis of dual-hop af systems with interference in
  nakagami-$m$ fading channels,'' \emph{IEEE Signal Processing Letters},
  vol.~18, no.~8, pp. 454 --457, Aug. 2011.

\bibitem{SMY:12:TVT}
H.~A. Suraweera, D.~S. Michalopoulos, and C.~Yuen, ``Performance analysis of
  fixed gain relay systems with a single interferer in nakagami-$m$ fading
  channels,'' \emph{IEEE Transactions on Vehicular Technology}, vol.~61, no.~3,
  pp. 1457 --1463, Mar. 2012.

\bibitem{IA:10:GLOBCOM}
S.~S. Ikki and S.~Aissa, ``Performance analysis of dual-hop relaying systems in
  the presence of co-channel interference,'' in \emph{Proc. IEEE Global
  Telecommunications Conference, GLOBECOM 2010}, Dec. 2010, pp. 1 --5.

\bibitem{CDS:11:IET}
A.~M. Cvetkovic´, G.~T. Ðordevic´, and M.~C. Stefanovic´, ``Performance of
  interference-limited dual-hop non-regenerative relays over rayleigh fading
  channels,'' \emph{IET Communications}, vol.~5, no.~2, pp. 135--140, 21 2011.

\bibitem{XZZ:11:CL}
W.~Xu, J.~Zhang, and P.~Zhang, ``Outage probability of two-hop fixed-gain relay
  with interference at the relay and destination,'' \emph{IEEE Communications
  Letters}, vol.~15, no.~6, pp. 608 --610, Jun. 2011.

\bibitem{LL:11:TVT}
D.~Lee and J.~H. Lee, ``Outage probability for dual-hop relaying systems with
  multiple interferers over rayleigh fading channels,'' \emph{IEEE Transactions
  on Vehicular Technology}, vol.~60, no.~1, pp. 333 --338, Jan. 2011.

\bibitem{CY:11:CL}
D.~B. da~Costa and M.~D. Yacoub, ``Outage performance of two hop af relaying
  systems with co-channel interferers over nakagami-m fading,'' \emph{IEEE
  Communications Letters}, vol.~15, no.~9, pp. 980 --982, Sept. 2011.

\bibitem{CZLY:10:VTC}
S.~Chen, X.~Zhang, F.~Liu, and D.~Yang, ``Outage performance of dual-hop relay
  network with co-channel interference,'' in \emph{Proc IEEE 71st Vehicular
  Technology Conference, VTC 2010-Spring}, May 2010, pp. 1 --5.

\bibitem{ZJW:10:TC}
C.~Zhong, S.~Jin, and K.-K. Wong, ``Dual-hop systems with noisy relay and
  interference-limited destination,'' \emph{IEEE Transactions on
  Communications}, vol.~58, no.~3, pp. 764--768, Mar. 2010.

\bibitem{CDG:11:CL}
D.~B. da~Costa, H.~Ding, and J.~Ge, ``Interference-limited relaying
  transmissions in dual-hop cooperative networks over nakagami-$m$ fading,''
  \emph{IEEE Communications Letters}, vol.~15, no.~5, pp. 503 --505, May 2011.

\bibitem{MDFT:11:IT}
S.~Mohajer, S.~N. Diggavi, C.~Fragouli, and D.~N.~C. Tse, ``Approximate
  capacity of a class of gaussian interference-relay networks,'' \emph{IEEE
  Transactions on Information Theory}, vol.~57, no.~5, pp. 2837--2864, may
  2011.

\bibitem{RDH:06:IZS}
T.~Riihonen, K.~Doppler, and A.~Hottinen, ``Performance of spatially
  distributed large interference relay networks,'' in \emph{Proc. International
  Zurich Seminar on Communications}, 0-0 2006, pp. 190--193.

\bibitem{MDFT:08:CCC}
S.~Mohajer, S.~N. Diggavi, C.~Fragouli, and D.~Tse, ``Transmission techniques
  for relay-interference networks,'' in \emph{Proc. 46th Annual Allerton
  Conference on Communication, Control, and Computing}, sept. 2008, pp.
  467--474.

\bibitem{TN:12:IT}
R.~A. Tannious and A.~Nosratinia, ``Relay-assisted interference network:
  Degrees of freedom,'' \emph{IEEE Transactions on Information Theory},
  vol.~PP, no.~99, p.~1, 2012.

\bibitem{ZB:05:CL}
X.~Zhang and N.~C. Beaulieu, ``Outage probability of mrc with equi-power
  cochannel interferers in correlated rayleigh fading,'' \emph{IEEE
  Communications Letters}, vol.~9, no.~7, pp. 577--579, july 2005.

\bibitem{ZB:06:CL}
------, ``Outage probability of mrc with unequal-power cochannel interferers in
  correlated rayleigh fading,'' \emph{IEEE Communications Letters}, vol.~10,
  no.~1, pp. 7 -- 9, jan 2006.

\bibitem{CZF:04:CL}
X.~W. Cui, Q.~T. Zhang, and Z.~M. Feng, ``Outage performance for maximal ratio
  combiner in the presence of unequal-power co-channel interferers,''
  \emph{IEEE Communications Letters}, vol.~8, no.~5, pp. 289 -- 291, may 2004.

\bibitem{CVA:03:TWC}
C.~Chayawan and V.~A. Aalo, ``Average error probability of digital cellular
  radio systems using mrc diversity in the presence of multiple interferers,''
  \emph{IEEE Transactions on Wireless Communications}, vol.~2, no.~5, pp.
  860--864, sept. 2003.

\bibitem{AZ:01:TVT}
V.~A. Aalo and J.~Zhang, ``Performance analysis of maximal ratio combining in
  the presence of multiple equal-power cochannel interferers in a nakagami
  fading channel,'' \emph{IEEE Transactions on Vehicular Technology}, vol.~50,
  no.~2, pp. 497--503, mar 2001.

\bibitem{JW:84:JSAC}
J.~Winters, ``Optimum combining in digital mobile radio with cochannel
  interference,'' \emph{IEEE Journal on Selected Areas in Communications},
  vol.~2, no.~4, pp. 528 --539, july 1984.

\bibitem{SH:98:TC}
A.~Shah and A.~M. Haimovich, ``Performance analysis of optimum combining in
  wireless communications with rayleigh fading and cochannel interference,''
  \emph{IEEE Transactions on Communications}, vol.~46, no.~4, pp. 473--479, apr
  1998.

\bibitem{VZ:00:CL}
V.~A. Aalo and J.~Zhang, ``Performance of antenna array systems with optimum
  combining in a rayleigh fading environment,'' \emph{IEEE Communications
  Letters}, vol.~4, no.~12, pp. 387--389, dec 2000.

\bibitem{CWZ:03:TC}
M.~Chiani, M.~Z. Win, and A.~Zanella, ``Error probability for optimum combining
  of m-ary psk signals in the presence of interference and noise,'' \emph{IEEE
  Transactions on Communications}, vol.~51, no.~11, pp. 1949 -- 1957, nov.
  2003.

\bibitem{LJJ:11:TWC}
L.~Li, Y.~Jing, and H.~Jafarkhani, ``Interference cancellation at the relay for
  multi-user wireless cooperative networks,'' \emph{IEEE Transactions on
  Wireless Communications}, vol.~10, no.~3, pp. 930--939, march 2011.

\bibitem{VL:08:TWC}
S.~Vakil and B.~Liang, ``Cooperative diversity in interference limited wireless
  networks,'' \emph{IEEE Transactions on Wireless Communications}, vol.~7,
  no.~8, pp. 3185--3195, august 2008.

\bibitem{SW:08:IT}
H.~Shin and M.~Z. Win, ``{MIMO} diversity in the presence of double
  scattering,'' \emph{IEEE Transactions on Information Theory}, vol.~54, no.~7,
  pp. 2976--2996, Jul. 2008.

\bibitem{BSW:07:WCOM}
A.~Bletsas, H.~Shin, and M.~Z. Win, ``Cooperative communications with
  outage-optimal opportunistic relaying,'' \emph{IEEE Transactions on Wireless
  Communications}, vol.~6, no.~9, pp. 3450--3460, Sep. 2007.

\bibitem{PBM:90:Book:v3}
A.~P. Prudnikov, Y.~A. Brychkov, and O.~I. Marichev, \emph{Integrals and
  Series}.\hskip 1em plus 0.5em minus 0.4em\relax New York: Gordon and Breach
  Science, 1990, vol.~3.

\bibitem{GR:07:Book}
I.~S. Gradshteyn and I.~M. Ryzhik, \emph{Table of Integrals, Series, and
  Products}, 7th~ed.\hskip 1em plus 0.5em minus 0.4em\relax San Diego, CA:
  Academic, 2007.

\bibitem{GDC:09:TWC}
B.~Gui, L.~Dai, and L.~Cimini, ``Routing strategies in multihop cooperative
  networks,'' \emph{IEEE Transactions on Wireless Communications}, vol.~8,
  no.~2, pp. 843--855, Feb. 2009.

\bibitem{PBM:92:Book:v4}
A.~P. Prudnikov, Y.~A. Brychkov, and O.~I. Marichev, \emph{Integrals and
  Series: Direct Laplace Transforms}.\hskip 1em plus 0.5em minus 0.4em\relax
  New York: Gordon and Breach Science, 1992, vol.~4.

\end{thebibliography}

\newpage


\end{document}